\documentclass[11pt]{article}
\usepackage{CJKnumb}
\usepackage{amsmath, amsfonts, bm,cite}
\usepackage{newtxtext} 
\usepackage{newtxmath} 
\usepackage[pdftex]{graphicx}
\usepackage{color}
\usepackage{multirow}
\usepackage{fancyhdr}
\usepackage{array}
\usepackage{makeidx, chngpage}
\usepackage{enumerate}
\usepackage{titlesec}
\usepackage[T1]{fontenc}
\usepackage{float}
\usepackage{url}
\usepackage{caption}
\usepackage{booktabs}
\usepackage{ascii}
\usepackage[margin=1.5cm]{geometry}
\usepackage{bm}
\usepackage{indentfirst}
\usepackage{textcomp}
\usepackage{graphicx}
\usepackage{hyperref} 
\hypersetup{
    colorlinks=true,        
    linkcolor=blue,         
    citecolor=blue,          
    urlcolor=purple         
}
\usepackage{mwe}
\usepackage{subfigure}
\usepackage{appendix}
\usepackage{mathtools}

\usepackage{amstext}
\usepackage{fancyhdr}
\usepackage{lineno}
\nolinenumbers

\RequirePackage[table]{xcolor} 
\RequirePackage{soul} 
\RequirePackage{multirow}

\newlength{\defbaselineskip}
\newcommand{\bitem}{\begin{itemize}} 
\newcommand{\eitem}{\end{itemize}}
\newcommand{\be}{\begin{equation}}
\newcommand{\ee}{\end{equation}}
\newcommand{\benu} {\begin{enumerate}}
\newcommand{\eenu} {\end{enumerate}}
\newcommand{\ba}{\begin{eqnarray}}
\newcommand{\ea}{\end{eqnarray}}
\newcommand{\bdes}{\begin{description}}
\newcommand{\edes}{\end{description}}

\title{Uncertainty-Aware Digital Twins: Robust Model Predictive Control using Time-Series Deep Quantile Learning}
\author{Yi-Ping Chen$^{1}$, Ying-Kuan Tsai$^{1}$, Vispi Karkaria$^{1}$, and Wei Chen$^{1}$ \thanks{Corresponding author, weichen@northwestern.edu, Department of Mechanical Engineering, Northwestern University, Evanston, IL, 60208}\\ \small{$^{1}$ Department of Mechanical Engineering, Northwestern University}} 
\date{}

\begin{document}

\graphicspath{{Graphs/}}

\maketitle

\begin{abstract}
Digital Twins, virtual replicas of physical systems that enable real-time monitoring, model updates, predictions, and decision-making, present novel avenues for proactive control strategies for autonomous systems. However, achieving real-time decision-making in Digital Twins considering uncertainty necessitates an efficient uncertainty quantification (UQ) approach and optimization driven by accurate predictions of system behaviors, which remains a challenge for learning-based methods. This paper presents a simultaneous multi-step robust model predictive control (MPC) framework that incorporates real-time decision-making with uncertainty awareness for Digital Twin systems. Leveraging a multi-step-ahead predictor named Time-Series Dense Encoder (TiDE) as the surrogate model, this framework differs from conventional MPC models that provide only one-step ahead predictions. In contrast, TiDE can predict future states within the prediction horizon in one-shot, significantly accelerating MPC. Furthermore, quantile regression is employed with the training of TiDE to perform flexible while computationally efficient UQ on data uncertainty. Consequently, with the deep learning quantiles, the robust MPC problem is formulated into a deterministic optimization problem and provides a safety buffer that accommodates disturbances to enhance constraint satisfaction rate. As a result, the proposed method outperforms existing robust MPC methods by providing less-conservative UQ and has demonstrated efficacy in an engineering case study involving Directed Energy Deposition (DED) additive manufacturing. This proactive while uncertainty-aware control capability positions the proposed method as a potent tool for future Digital Twin applications and real-time process control in engineering systems.
\end{abstract}

\section{Introduction}

\subsection{Problem Definition}

The concept of Digital Twins \cite{national2023foundational, van2023digital} has shown promising revolutions in autonomous industries such as manufacturing~\cite{karkaria2024towards,karkaria2025optimization,zemskov2024security} and predictive maintenance~\cite{zhong2023overview}. It brings the idea of building bi-directional interactions between the physical system and its virtual counterpart.  This enables online decision-making processes to be conducted automatically utilizing the state prediction provided by the virtual systems, and reacts proactively in response to the feedback from the physical systems \cite{grieves2017digital}. One embodiment of online decision-making for Digital Twins is via model predictive control (MPC) \cite{rawlings2017model}, which optimizes system performance by predicting future behavior and adjusting control inputs in real-time based on the model prediction. To account for disturbances in MPC, a family of uncertainty-aware MPC approaches has been proposed to enhance constraint satisfaction rates in the presence of anticipated uncertainty. Methods such as stochastic MPC \cite{mesbah2016stochastic} and robust MPC \cite{mayne2005robust}, aim to approximate uncertainty propagation through the known system dynamics, quantify the distribution of predicted states, and solve the MPC problem by explicitly incorporating uncertainty bounds with the constraints. As the accurate description of the system dynamics may be unavailable a priori, with the recent advancements in machine learning and neural networks (NN) \cite{karkaria2025optimization}, learning-based or data-driven predictive controllers \cite{manzano2020robust, hewing2020learning} have gained significant attention. However, although NNs can emulate the system dynamics accurately, applying NNs in uncertainty-aware MPC presents significant challenges, as quantifying and estimating uncertainty distributions can be both complex and computationally expensive. Consequently, integrating NN-based models with efficient uncertainty quantification (UQ) methods for decision-making remains an open research question critical to the advancement of Digital Twin technologies.

\subsection{Uncertainty-Aware MPC}

Uncertainty-aware MPC methods can be broadly categorized into robust and stochastic approaches~\cite{rawlings2017model}. Robust MPC focuses on optimizing control inputs to perform effectively under worst-case scenarios, ensuring system stability and constraint satisfaction even under bounded uncertainties~\cite{mayne2005robust}. Stochastic MPC leverages probabilistic models to incorporate uncertainties into the optimization process and formulate probabilistic (or chance) constraints in an optimal control problem~\cite{mesbah2016stochastic}, aiming to achieve a balance between performance and reliability. Techniques like min-max MPC formulates the cost function as the maximum of cost values with the samples generated based on disturbance models~\cite{lofberg2003minimax}. However, min-max MPC often comes with significant computational overhead, posing challenges for real-time implementation~\cite{bemporad2007robust,gonzalez2011online,tsai2023robust}. To make these optimal control problems more computationally tractable, tube-based techniques have been explored to solve the robust and stochastic MPC problems by explicitly identifying the uncertainty regions in state and control action spaces~\cite{langson2004robust,lorenzen2016constraint,tsai2023robust,tsai2025control}. However, many of these approaches still assume prior knowledge of system dynamics or disturbance characteristics, limiting their applicability in real-world scenarios with incomplete or evolving information. In recent years, data-driven modeling, which can capture complex temporal dependencies and non-stationary dynamics, has gained attention for MPC frameworks. By leveraging these advances, the integration of data-driven methods into MPC frameworks offers new opportunities to improve both performance and adaptability under uncertain conditions.


\subsection{Data-Driven MPC}
Data-driven models are essential for surrogating physics in a Digital Twin and MPC, particularly in two key scenarios: when the system's underlying physics is overly complex or not fully understood, and when simulations are prohibitively computationally expensive or time-consuming \cite{Thelen2022}. Under these circumstances, data-driven/learning-based methods can identify the system directly using observational data. For example, neural state space models \cite{drgovna2022differentiable} can replace the system and input matrices in a state space formulation. Recursive Neural Network (RNN) and Long Short-Term Memory (LSTM) are also popular options since their structures resemble the propagation of the dynamics of the systems \cite{jung2023model, wu2019real, huang2022lstm}. However, enabling learning-based methods with uncertainty-awareness for real-time applications is still challenging, primarily due to the computational complexity of performing UQ. Popular UQ techniques for NNs, such as ensemble methods, Bayesian NNs, Monte Carlo (MC) dropout, and bootstrapping, fall under the category of sampling-based methods \cite{kabir2018neural}. While these methods can numerically approximate the distribution of NN outputs \cite{ren2024advancing}, their reliance on Monte Carlo sampling and multiple forward passes of NNs suffers from significant computational time, rendering them impractical for many engineering applications involving MPC.

\begin{figure}[b]
    \centering
    \includegraphics[width=1\linewidth]{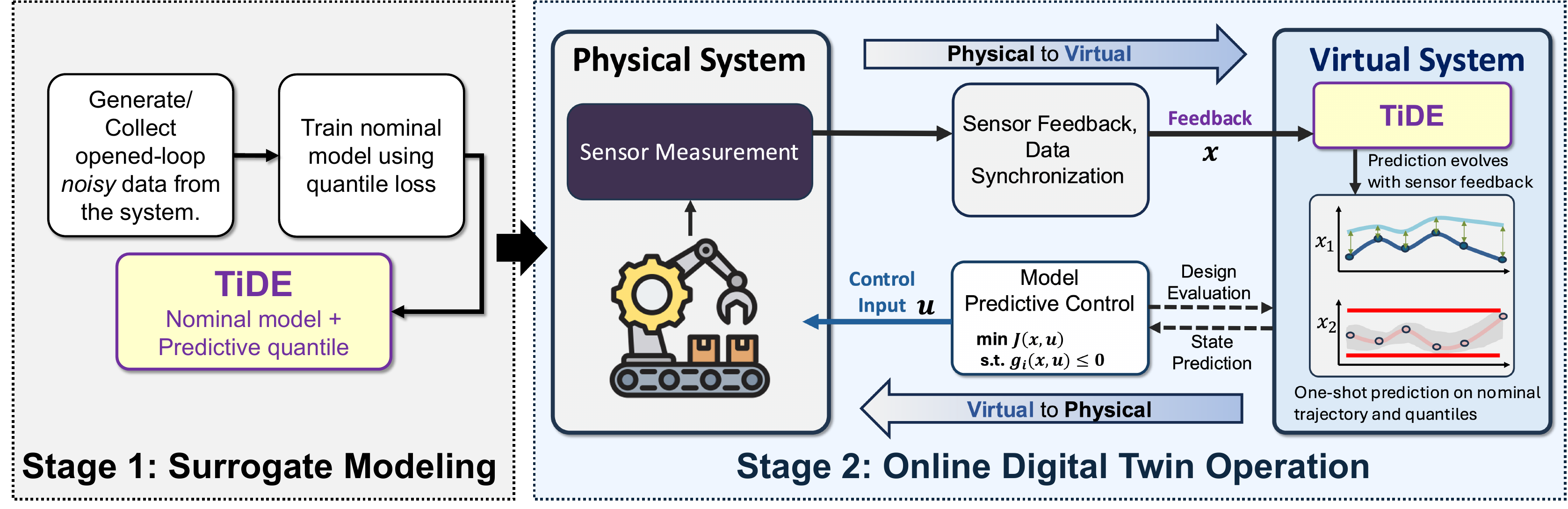}
    \caption{Proposed multi-step robust MPC framework.}
    \label{fig:overall_flowchart}
\end{figure}

In contrast, parametric methods, which estimate the parameters of the uncertainty distribution directly, provide a computationally efficient alternative and are widely applied in learning-based MPC. For instance, Kinky Inference has been employed to learn parameters representing the bounds of system states \cite{manzano2020robust}. Similarly, in \cite{hewing2019cautious}, a Gaussian process (GP) is utilized as a discrepancy model to capture unknown system dynamics, with its predictive uncertainty serving as a probabilistic bound for nominal predictions. One noteworthy approach is quantile regression, which directly learns user-defined quantiles of the data. Unlike methods that require assumptions about the data distribution, quantile regression offers greater flexibility, making it particularly appealing for MPC application \cite{fan2020deep}.

Another challenge in data-driven MPC lies in the significant computational cost of solving optimization problems online. Unlike linear MPC, which benefits from efficient closed-form or QP-based solutions via linear quadratic regulation (LQR) \cite{rawlings2017model}, data-driven MPC relies on surrogate models, e.g. black-box functions such as neural networks, which require iterative numerical solvers and multiple forward evaluations. This becomes particularly burdensome when using one-step-ahead predictors, which must be rolled out recursively to generate the full trajectory prediction. 

\subsection{MPC with Multi-Step-Ahead Predictors}

To alleviate the computational burden, a growing body of research has turned toward multi-step-ahead predictors, taking advantage of the recent advance in machine learning. These models generate the entire future trajectory in a single forward pass, reducing the number of function evaluations during optimization and thus significantly lowering computational cost. For example, Park \cite{park2023simultaneous} demonstrated that using multi-step predictors based on Transformer architectures yielded substantial improvements in runtime and prediction accuracy, especially for longer horizons. This benefit is particularly attractive in real-time MPC settings, where computational efficiency is critical.

Moreover, multi-step predictors provide structural advantages for UQ \cite{kohler2022state, terzi2019learning}. Traditional robust MPC approaches, such as tube-based methods, typically require recursive uncertainty propagation, which can lead to overly conservative UQ. Recent work, such as \cite{kohler2022state}, highlights that multi-step predictors allow uncertainty to be learned directly from data at the trajectory level, bypassing the need for recursive propagation and enabling simpler, more intuitive bounds. This is especially valuable in data-driven contexts, where model structure may not be fully known, and where robust decision-making must account for uncertainty without sacrificing computational tractability. Uncertainty estimation in existing methods for robust MPC with multi-step-ahead predictors \cite{kohler2022state, terzi2022robust} focusing on deriving the worst-case using linear models with Set Membership identification. Although they  provide intepretability and theoretical rigors in estimating erro bound, these approaches are limited in linear models and can hardly be generalized to NN-based applications.

In parallel, advances in time-series forecasting have demonstrated the effectiveness of quantile regression integrated with sequence-to-sequence deep learning models for multi-step uncertainty estimation \cite{cheung2024quantile, hu2024novel}. These models can capture complex temporal patterns and provide prediction intervals across time, offering a natural fit for robust decision-making that require trajectory-wise uncertainty estimates. Taken together, these trends point to multi-step-ahead prediction and direct UQ learning such as quantile regression as promising directions for improving both efficiency and robustness in data-driven MPC, particularly under real-time and uncertain environments in Digital Twin applications.

\subsection{Research Objective}
This work introduces a simultaneous multi-step robust MPC framework that leverages time-series deep neural networks and deep quantile learning to enable fast, uncertainty-aware decision-making in Digital Twins of complex engineering systems. While previous works have explored multi-step predictions to accelerate MPC \cite{park2023simultaneous} and employed quantile regression to quantify uncertainty in single-step MPC \cite{fan2020deep}, this work is, to the best of our knowledge, the first to unify these two paradigms, enabling deep learning quantiles to be explicitly used as predicted bounds in multi-step robust MPC setting. By combining the computational efficiency of multi-step-ahead predictors with the flexibility and accuracy of trajectory-level uncertainty quantification, our framework opens a new direction for robust MPC that is both data-driven and scalable. As MPC is considered a popular model-based decision-making approach that can be built on existing Digital Twin frameworks, such as \cite{karkaria2025optimization,karkaria2024towards,thelen2022comprehensive, national2023foundational}, this proposed method pushes the limit of current MPC, providing a new option for decision-making under uncertainty in Digital Twin applications.

The proposed framework, depicted in Fig. \ref{fig:overall_flowchart}, comprises two stages. In Stage 1, noisy system data is gathered as the training data. A time-series deep neural network, named Time-Series Dense Encoder (TiDE), is employed to perform nonlinear system identification, capturing both the nominal system dynamics and the quantiles of the data uncertainty, encompassing the uncertainty of the nominal prediction. Subsequently, in Stage 2, TiDE serves as the predictive model (virtual system) operating with the proposed multi-step robust MPC as the virtual-to-physical integration. The nominal prediction (median) is utilized to assess the reference tracking performance, while the predictive quantiles are employed to guarantee constraint satisfaction. This proposed method is validated using an illustrative example and an engineering case study in additive manufacturing. The contributions of this work include:

\begin{itemize}
    \item We propose a robust MPC framework for multi-step ahead prediction models as an embodiment of uncertainty-aware real-time decision-making for Digital Twins.
    
    \item We demonstrate the effectiveness of deep learning quantiles in quantifying data uncertainty.

    \item We validate the proposed methods using several case studies, showing the generality of this method in the Digital Twin paradigm. 
\end{itemize}

The rest of the paper is structured as follows: In Section 2, the technical background of MPC, TiDE, and quantile regression will be introduced. Section 3 details the proposed robust MPC framework, including problem formulation, model preparation, and optimization techniques. In Section 4, a numerical model is used as a demonstration to walk through the implementation details, and the result in an engineering case study on additive manufacturing (AM) is revealed in Section 5. Lastly, we will conclude this work in Section 6.

\section{Technical Background}

\emph{Notation:} The sets of real numbers and non-negative integers are denoted by $\mathbb{R}$ and $\mathbb{N}_{\geq0}$, respectively. Given $a,b\in\mathbb{N}_{\geq0}$ such that $a<b$, we denote $\mathbb{N}_{[a,b]}:=\{a,a+1,...,b\}$. $[\mathbf{A}]_i$ and $[\mathbf{a}]_i$ denote the $i$th row and element of the matrix $\mathbf{A}$ and vector $\mathbf{a}$, respectively. $\hat{\mathbf{x}}_{k+i}$ denotes the $i$-step-ahead predicted value of $\mathbf{x}$ at time $k$. The notation $\mathbf{I}_{a\times a}$ denotes an $a$-by-$a$ identity matrix, $\mathbf{Q}\succ0$ indicates a positive definite matrix, and $||\mathbf{x}||_{\mathbf{Q}}^2=\mathbf{x}^\top\mathbf{Q}\mathbf{x}$ refers to a quadratically weighted norm. Given a random variable $X$, $\mathbb{E}[X]$ denotes its expected value. A Gaussian distribution with mean vector $\bm\mu$ and covariance matrix $\mathbf{\Sigma}$ is represented as $\mathcal{N}(\bm\mu,\mathbf{\Sigma})$. Given two sets $A$ and $B$, then $A\oplus B:=\{a+b|a\in A,~b\in B\}$ (Minkowski sum) and $A\ominus B:=\{a\in A|a+b\in A, \forall b\in B\}$ (Pontryagin difference).

\subsection{MPC and Robust MPC}
\begin{figure}[h!]
    \centering
    \includegraphics[width=0.6\linewidth]{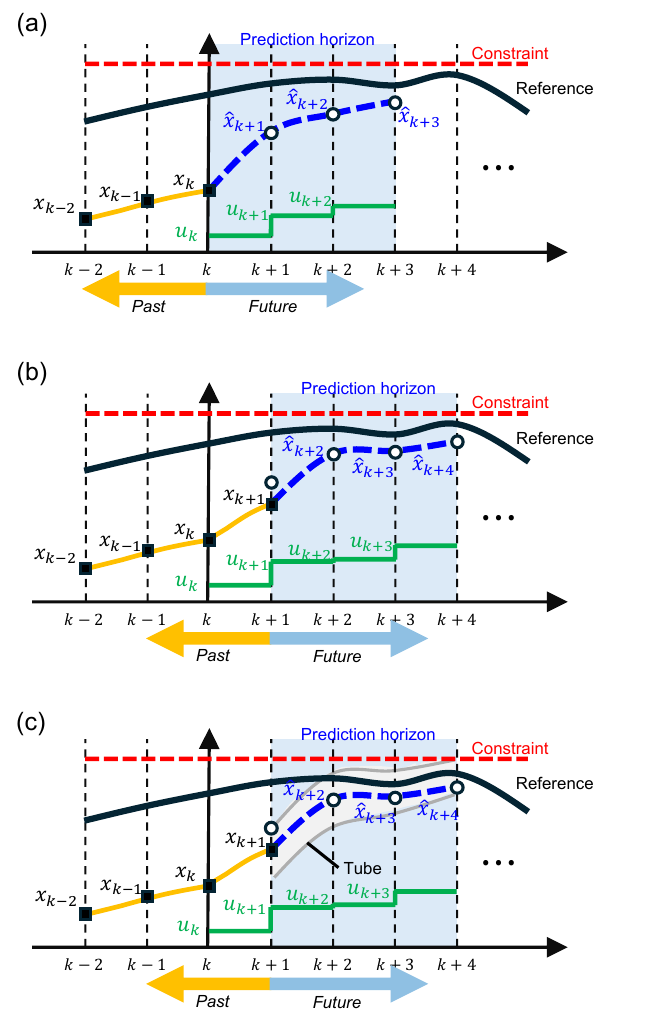}
    \caption{Illustration of MPC and robust MPC. (a) Illustration of MPC at time = $k$. (b) Illustration of MPC at time $k+1$. (c) Illustration of robust MPC. The green line are the optimal control input sequences, and the blue dash lines are the state prediction from the model given the optimal control inputs. The gray tube in (c) represents the quantified uncertainty. }
    \label{fig:MPC_illustration}
\end{figure}

Model Predictive Control (MPC), as known as receding horizon optimal control, is an advanced control method that employs an explicit dynamic model of the system to predict and optimize future control actions within a finite horizon \cite{rawlings2017model}. This ensures that constraints on inputs and outputs are met while minimizing a specified cost function. MPC iteratively solves an optimization problem at each time step, applies the resulting control action, and repeats the process as the time horizon advances, as depicted in Fig. \ref{fig:MPC_illustration}(a) and Fig. \ref{fig:MPC_illustration}(b).

Assume that a general nonlinear system can be represented as:
\begin{equation}
    \mathbf{x}_{k+1} = F(\mathbf{x}_k,\mathbf{u}_{k}),~\forall k\in\mathbb{N},
    \label{eq:dynamic_eq}
\end{equation}
where $F$ denotes the dynamic function that maps from the current state and control action to the state at the next step. With the prediction horizon length noted as $N$ and the specified reference $\mathbf{r}=[\mathbf{r}_{k+1},...,\mathbf{r}_{k+N}]$, the MPC can be formulated as an optimization problem for the future control inputs $\mathbf{u}=[\mathbf{u}_k,...,\mathbf{u}_{k+N-1}]$:
\begin{subequations}
\begin{align}
    \min_{\mathbf{u}} ~&J(\mathbf{u},\hat{\mathbf{x}},\mathbf{r})=\sum_{i=0}^{N-1}\left[||\hat{\mathbf{x}}_{k+i+1}-\mathbf{r}_{k+i+1}||_{\mathbf{Q}}^2+||\mathbf{u}_{k+i}||_{\mathbf{R}}^2\right], \label{eq:MPC_objective} \\
    s.t. \quad & \hat{\mathbf{x}}_{k+1}=\hat{F}(\mathbf{x}_{k},\mathbf{u}_{k}),\label{eq:MPC_constraints_m} \\
    & \hat{\mathbf{x}}_{k+i+1}=\hat{F}(\hat{\mathbf{x}}_{k+i},\mathbf{u}_{k+i}),~\forall i\in\mathbb{N}_{[1,N-1]},\label{eq:MPC_constraints_b}\\
       & \hat{\mathbf{x}}_{k+i}\in\mathbb{X},~\forall i\in\mathbb{N}_{[1,N]},\label{eq:MPC_constraints_c}\\
      &  \mathbf{u}_{k+i}\in\mathbb{U},~\forall i\in\mathbb{N}_{[0,N-1]},\label{eq:MPC_constraints_d} \\
      & \mathbf{g}(\hat{\mathbf{x}}_{k+i+1},\mathbf{u}_{k+i}) \leq \mathbf{0}, ~\forall i\in\mathbb{N}_{[0,N-1]}, \label{eq: MPC_constraint_e} \\
      & \mathbf{h}(\hat{\mathbf{x}}_{k+i+1},\mathbf{u}_{k+i}) = \mathbf{0}, ~\forall i\in\mathbb{N}_{[0,N-1]}, \label{eq: MPC_constraint_f}
    \end{align}
    \label{eq:MPC_constraints}
\end{subequations} 

\noindent where $\|\mathbf{x}\|_{\mathbf{Q}}^2=\mathbf{x}^\top\mathbf{Q}\mathbf{x}$ represents the quadratic operation on the state vector $\mathbf{x}$, the weighting matrices $\mathbf{Q}\succ0$ and $\mathbf{R}\succ0$ are symmetric. The MPC objective function $J$ in Eq. (\ref{eq:MPC_objective}) consists of two type of loss: $||\hat{\mathbf{x}}_{k+i+1}-\mathbf{r}_{k+i+1}||_{\mathbf{Q}}^2$ refers to the reference tracking error throughout the horizon using the predicted states, and $||\mathbf{u}_{k+i}||_{\mathbf{R}}^2$ penalizes the control efforts. The two loss terms can be balanced by user-selected $\mathbf{Q}$ and $\mathbf{R}$. Eq. (\ref{eq:MPC_constraints_b}) is the general representation of the dynamic equation, where $\hat{F}$ denotes the predictive model. Eq. (\ref{eq:MPC_constraints_c}) and (\ref{eq:MPC_constraints_d}) are the constraints on states and control actions, respectively, while Equations (\ref{eq: MPC_constraint_e}) and (\ref{eq: MPC_constraint_f}) explicitly denote all the inequality and equality constraints if any applies.

While traditional MPC effectively optimizes control actions within a finite horizon based on a deterministic system model~\cite{tsai2022design}, it does not inherently account for uncertainties or disturbances that can impact the system dynamics, states, or constraints. The general nonlinear system with uncertainties can be represented as:
\begin{equation}
\mathbf{x}_{k+1} = {F}_w(\mathbf{x}_k, \mathbf{u}_k, \mathbf{w}_k),
\label{eq:dynamic_eq_uncertainty}
\end{equation}
where $\mathbf{w}_k$ represents the disturbance vector, often assumed to lie within a known set $\mathbb{W}$ or be independent and identically normally distributed with zero means and a diagonal covariance matrix $\boldsymbol{\Sigma}$:
\begin{equation}
\mathbf{w}_k \in \mathbb{W},~~\text{or}~~\mathbf{w}_k\sim \mathcal{N}(\mathbf{0},\mathbf{\Sigma_w}),
\label{eq:uncertainty_set}
\end{equation}
where $\mathbf{\Sigma_w}=\text{\rm diag}\left(\sigma^2_{w^{(1)}},...,\sigma^2_{w^{(n)}}\right)$.

To address this limitation, robust MPC extends the traditional MPC framework by explicitly incorporating uncertainties into the optimization problem, ensuring constraint satisfaction under the effect of uncertainties. The robust MPC optimization problem can be formulated by:
\begin{subequations}
\begin{align}
    \min\limits_{\mathbf{u}} ~&J(\mathbf{u},\hat{\mathbf{x}},\mathbf{r})=\sum_{i=0}^{N-1}\left[||\hat{\mathbf{x}}_{k+i+1}-\mathbf{r}_{k+i}||_{\mathbf{Q}}^2+||\mathbf{u}_{k+i}||_{\mathbf{R}}^2\right], \label{eq:RMPC_objective} \\
    s.t. \quad & \hat{\mathbf{x}}_{k+1}=\hat{F}_w(\mathbf{x}_{k},\mathbf{u}_{k}),\label{eq:RMPC_constraints_m}\\
    & \hat{\mathbf{x}}_{k+i+1}=\hat{F}_w(\mathbf{x}_{k+i},\mathbf{u}_{k+i}),~\forall i\in\mathbb{N}_{[0,N-1]},\label{eq:RMPC_constraints_a}\\
       & \hat{\mathbf{x}}_{k+i}\in\mathbb{X},~\forall i\in\mathbb{N}_{[0,N-1]},\label{eq:RMPC_constraints_b}\\
      &  \mathbf{u}_{k+i}\in\mathbb{U},~\forall i\in\mathbb{N}_{[0,N-1]},\label{eq:RMPC_constraints_c} \\
      & \mathbf{g}(\hat{\mathbf{x}}_{k+i+1},\mathbf{u}_{k+i}) \leq \mathbf{0}, ~\forall i\in\mathbb{N}_{[0,N-1]}, \label{eq: RMPC_constraint_d} \\
      & \mathbf{h}(\hat{\mathbf{x}}_{k+i+1},\mathbf{u}_{k+i}) = \mathbf{0}, ~\forall i\in\mathbb{N}_{[0,N-1]}. \label{eq: RMPC_constraint_e}
    \end{align}
    \label{eq:RMPC_constraints}
\end{subequations} 
Here, $\hat{F}_w$ is a surrogate model trained using noisy data. $\hat{\mathbf{x}}_k$ is a general representation of state prediction that can either be deterministic or stochastic. Note that since the disturbance is unknown to $\hat{F}_w$ when making state prediction, in contrast to Eq. \ref{eq:dynamic_eq_uncertainty}, $\mathbf{w}_k$ is not explicitly treated as the input of $\hat{F}_w$. 

Among all robust MPC techniques, min-max MPC is easy to implement because the solving procedure does not differ from conventional MPC. The major difference is that min-max MPC handles uncertainties by defining the cost function as the maximum of cost values over all realizations of disturbance sequences by multiple evaluations~\cite{lofberg2003minimax}. It is straightforward but inefficient because simulating all possible disturbances requires considerable cost and computational effort. Another drawback of min-max formulations is that the method results in too conservative solutions that restrict the operation and performance of the system~\cite{gonzalez2011online,zhang2013stochastic,9276487}. Although generating more samples to simulate disturbances can prevent such solutions, the online MPC computation becomes more time-consuming, leading to delayed system actuation. 

Tube-based MPC can be used to solve robust MPC problems by explicitly identifying the actual state region surrounding the nominal trajectory (called \emph{tube}), illustrated in Fig.~\ref{fig:MPC_illustration}(c). A tube accounts for deviations caused by uncertainties and can be included in the robust MPC formulation to satisfy the constraints for all realizations of disturbances~\cite{mayne2005robust,rawlings2017model}. However, tube-based MPC approaches typically rely on the assumptions of model representations (e.g., linear dynamic model~\cite{langson2004robust,lorenzen2016constraint,tsai2023robust,tsai2025control} or nonlinear model with Lipschitz functions~\cite{gao2014tube}) and disturbance types. Due to these assumptions, implementing tube-based MPC approaches to complex dynamic models presents a significant challenge.

\subsection{Time Series Deep Neural Network}

\begin{figure}[h!]
    \centering
\includegraphics[width=0.6\linewidth]{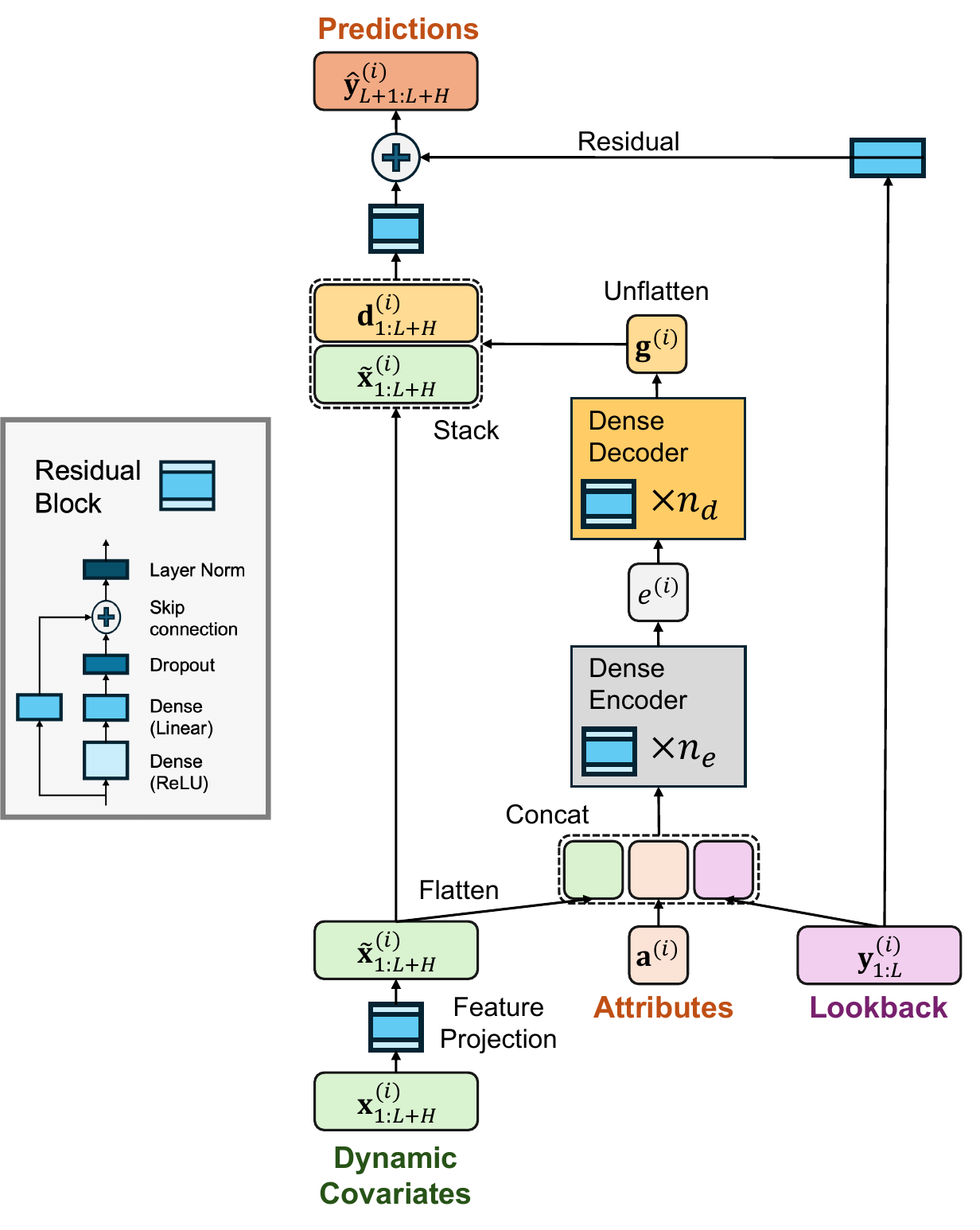}
    \caption{Network structure of TiDE, modified from \cite{das2023long}.}
    \label{fig:enter-label}
\end{figure}

There are two main considerations when selecting a suitable time-series DNN for MPC: 1) The inference speed should be fast as the solving process of MPC requires several function/model evaluations, and 2) The structure of the DNN should accommodate the general format of dynamical systems as denoted in Eq. (\ref{eq:dynamic_eq}). In this work, we select TiDE \cite{das2023long}, illustrated in Fig. \ref{fig:enter-label}, as the DNN for surrogating dynamical systems, due to its forward speed, model accuracy, and the compatibility of its input structure. TiDE's architecture, with parallelized dense layers and residual connections, ensures both computational efficiency and stable training for dynamic system modeling.

TiDE, designed with a residual network (ResNet) architecture for time-series data, leverages residual connections to enable effective gradient flow during backpropagation, preventing vanishing gradients and capturing long-term dependencies. Its reliance on dense layers allows it to process all time steps in parallel, making its forward pass faster than popular sequence-to-sequence models like Transformers and LSTM. Unlike LSTMs, which process sequences one step at a time recursively, TiDE operates on the entire input sequence as a batch, fully utilizing modern hardware like GPUs. While Transformers also parallelize, their self-attention introduces quadratic time complexity with respect to the sequence length, whereas TiDE's complexity grows linearly due to simple matrix multiplications. This linear complexity makes TiDE particularly suitable for real-time applications where low-latency predictions are critical. This efficiency allows TiDE to deliver faster inference while maintaining robust performance for time-series tasks.

The embedding capability of TiDE, realized by the dense encoder and decoders, improves predictions by transforming raw inputs into dense, low-dimensional representations that capture meaningful patterns and relationships. This reduces data dimensionality, encodes complex interactions, and enhances the model's ability to generalize across unseen examples. This dimensional-reduced embedding also plays the role as a noise filter by only identifying and embedding the most important features in its latent space. By effectively compressing input information, the embeddings help mitigate the risk of overfitting, especially in high-dimensional datasets. For time-series data, embeddings efficiently represent temporal attributes or categorical features, enabling TiDE to extract richer patterns and improve prediction accuracy.

\begin{figure}[h!]
    \centering
    \includegraphics[width=0.5\linewidth]{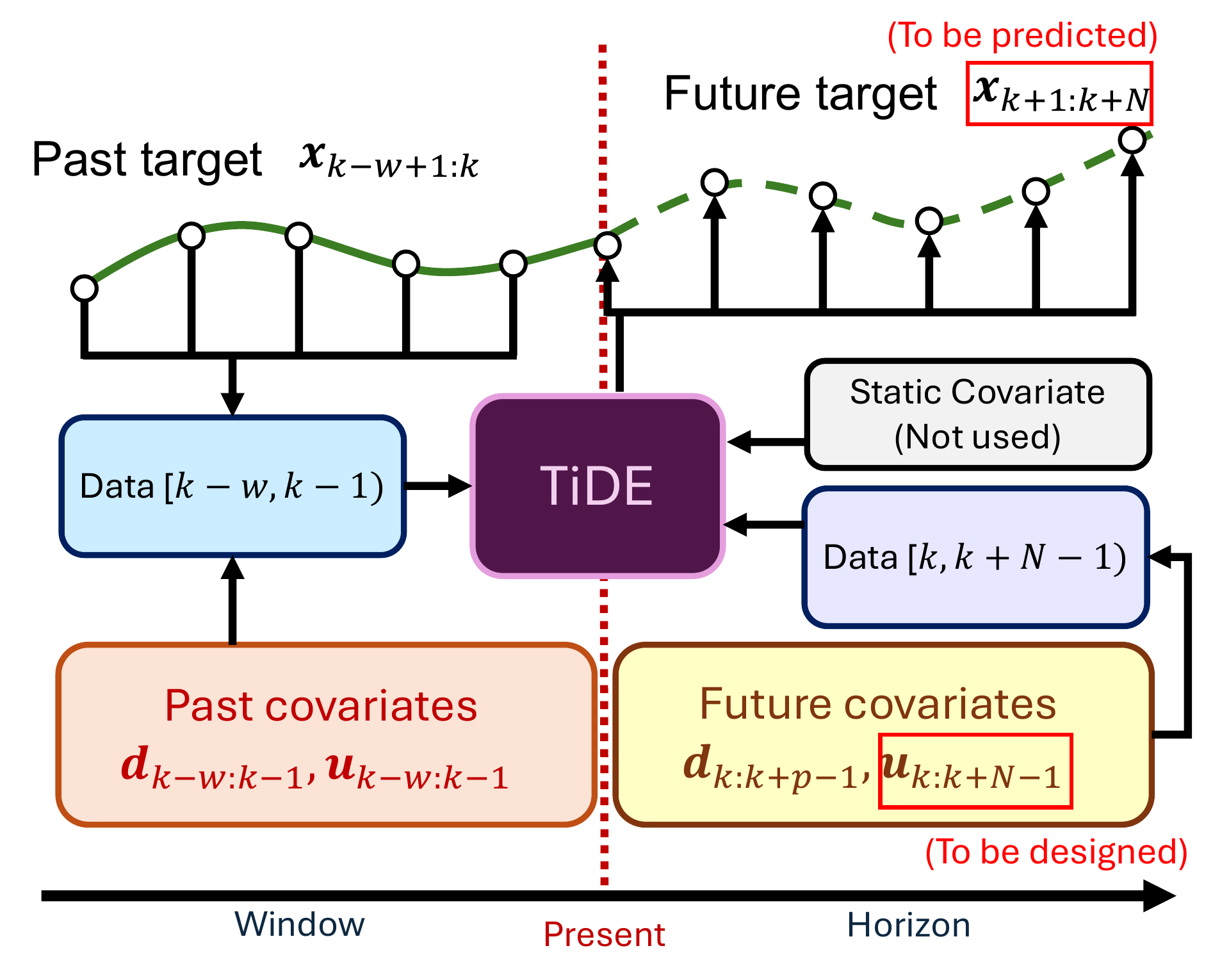}
    \caption{Data structure of the input and output of TiDE.}
    \label{fig:TiDE_data_structure}
\end{figure}

Different from conventional sequence-to-sequence prediction models, TiDE supports the usage of \emph{covariates} and \emph{targets} as the model input, as illustrated in Fig. \ref{fig:TiDE_data_structure}, making it suitable for surrogating dynamical systems. The target variable is the primary variable of interest in a time series forecasting model. It represents the value that is aimed to be predicted or forecasted, such as the future states of the system. The covariates are additional variables that provide supplementary information and can aid in predicting the target variable. They can be further categorized as past covariates and future covariates. These variables are often external or supplementary and are not part of the target series but are related to it. This structured separation of covariates and targets allows TiDE to capture both short-term dynamics and long-term dependencies more effectively. As shown in Fig. \ref{fig:TiDE_data_structure}, TiDE takes past target (e.g. past states $\mathbf{x}_{k-w+1:k}$, as an auto-regressive system), past covariates (e.g. past input $\mathbf{u}_{k-w:k-1}$ and other past input conditions $\mathbf{d}_{k-w:k-1}$ if required), and future covariates (e.g. future input $\mathbf{u}_{k:k+N-1}$ and other future input conditions $\mathbf{d}_{k:k+N-1}$) as model input to predict the future target $\mathbf{x}_{k+1:k+N}$ (future states). In particular, we denote $\mathbf{d}$ as the pre-defined system variables (e.g. the geometry information of a given part in additive manufacturing), and $\mathbf{u}$ as the future control input to be optimized in MPC (e.g. the laser power). Therefore, TiDE can be formulated as:
\begin{align}
    \hat{\mathbf{x}}_{k+1:k+N} = \text{TiDE}(\mathbf{x}_{k-w+1:k},& \mathbf{d}_{k-w:k-1}, \mathbf{u}_{k-w:k-1}, \mathbf{d}_{k:k+N-1},\mathbf{u}_{k:k+N-1} | \boldsymbol{\phi}), \label{eq:TiDE_general}
\end{align}
where $\boldsymbol{\phi}$ is the trainable NN parameters of TiDE. 

As a result, the separation of targets and covariates allows TiDE to resemble the nature of dynamical systems in a multi-step-ahead setting. In contrast, some forecasting models, such as Transformers \cite{park2023simultaneous} and N-BEATS \cite{oreshkin2019n}, predict future target solely based on the past target, but do not consider covariates explicitly. This design choice makes TiDE more versatile for applications where external influences, such as control inputs or environmental conditions, significantly affect system behavior.


\subsection{Quantile Regression}

Quantile regression (or quantile loss) \cite{koenker2001quantile, yu2003quantile, fan2020deep} is a versatile statistical technique used to estimate the conditional quantiles of a response variable, such as the median or other given percentiles, based on a set of predictor variables. Unlike ordinary least-squares (OLS) regression, which focuses on modeling the mean of the response variable, quantile regression captures a broader picture by modeling the entire conditional distribution. Specifically, quantile regression is effective in quantifying aleatoric uncertainty, which arises from inherent variability in the data. One key advantage of quantile regression is its robustness to outliers, as it is less sensitive to extreme values compared to methods such as mean square error (MSE) loss. This makes it particularly useful for datasets with skewed or irregular distributions. Moreover, quantile regression does not require prior assumptions about the distribution of the data, enabling it to handle heteroscedastic uncertainty, situations where the variability of the response changes across levels of the predictors. This flexibility allows the model to adapt to complex, real-world datasets where such variability is common.

The standard loss function for implementing quantile regression in supervised learning is defined as:
\begin{equation}
    L_{q,j}(y_t, \hat{y}_t) = 
\begin{cases} 
q \cdot (y_t - \hat{y}_t), & \text{if } y_t \geq \hat{y}_t, \\
(1 - q) \cdot (\hat{y}_t - y_t), & \text{if } y_t < \hat{y}_t.
\end{cases} \label{eq:standard_quantile}
\end{equation}
where the objective is to minimize the errors of a given quantile level $q$ (e.g. 0.5 for median) for response $j$. $y_t$ and $\hat{y}_t$ are the ground truth value and the predicted value of the target at time $t$, respectively. 

To justify the need of quantile regression, we perform a benchmark testing against popular UQ methods, including GP, MC dropout, ensemble methods, and deep evidential regression \cite{amini2020deep}. Here, a one-dimensional benchmark problem from \cite{amini2020deep} is modified by injecting non-Gaussian noise:
\begin{equation}
    y = x^3 + W_{exp}
    \label{eq:benchmark_problem}
\end{equation}
where $W_{exp}\sim \mathrm{Exp}(15)-15$ follows a zero mean Exponential distribution with rate $\lambda=15$. The reason for not assuming a normal distribution for the injected noise is that, even if the environmental disturbance is normally distributed, its propagation through a nonlinear system does not guarantee a normally distributed state. Therefore, when applying deep learning quantiles to nonlinear systems, assuming the state distribution is normal may not be the most appropriate. Aside from GP, all the NNs in each method have two hidden layers with size=64, and use ReLU() as the activation function, following the setting in \cite{amini2020deep}, while the output layer may be different, depending on each case. For training the model, 1334 training samples are generated within the range of $[-5,-4]\cup[-1,4]$, and all the NN models are trained with \texttt{Adam} for 500 epochs with learning rate $0.1$. To evaluate the methods, a test set with size 1000 is generated using $x\in[-7,7]$ and realized using Eq. \ref{eq:benchmark_problem}. Quantitative indices, including residual mean square error (RMSE), $R^2$, and coverage rate (num. of test samples fall within probabilistic bounds/total num. of test samples) are provided. Since the input range of training and test set are different, we not only evaluate the effecacy of learning noise distribution, but also to test the generalizability for extrapolation and interpolation at sparse data region.

\begin{table*}[t!]
\caption{Benchmarking of UQ methods. The lower figure to each method highlights the region within the pink box.} 
\label{table:UQ_benchmark}
\centering
\scriptsize
\begin{tabular}{c|c|c|c}
\hline \hline
     & \includegraphics[width=0.27\linewidth]{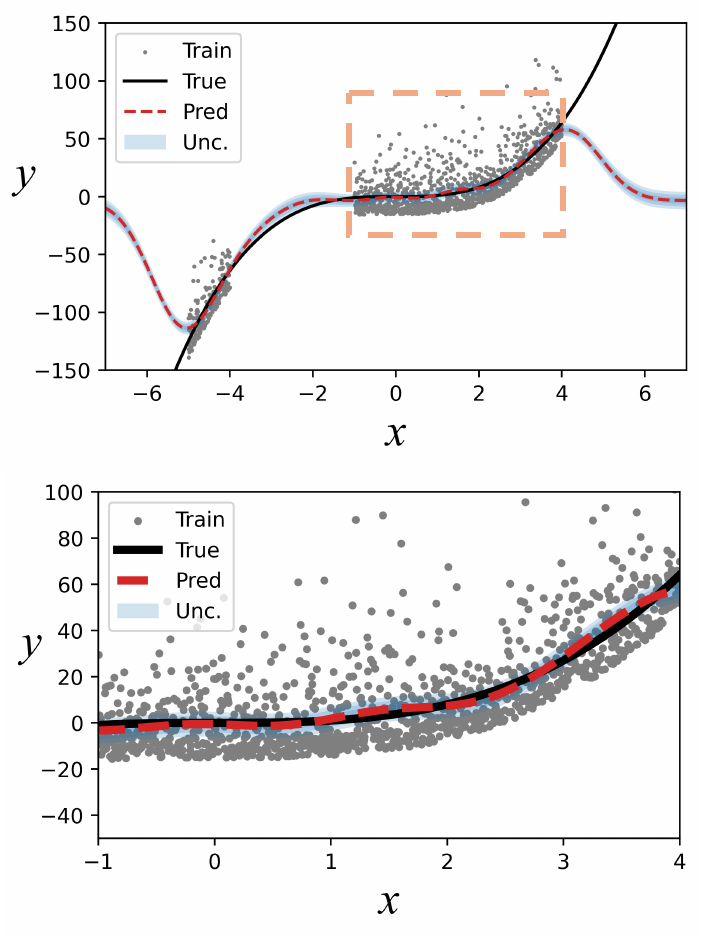}                                                        & \includegraphics[width=0.27\linewidth]{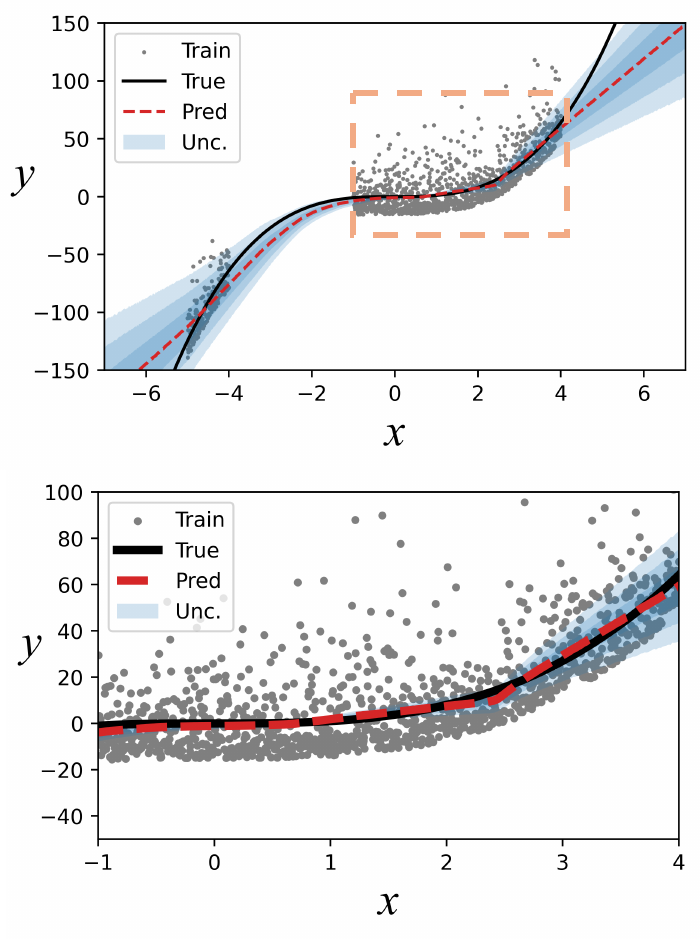}                                                                                       & \includegraphics[width=0.27\linewidth]{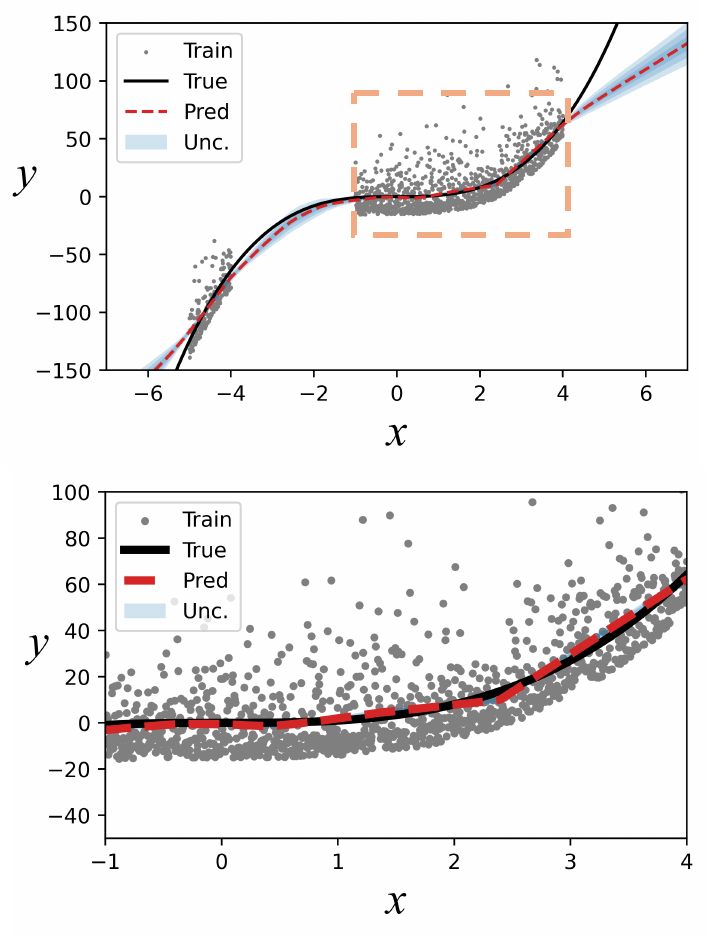}                 \\ 
Method                                                                  & Gaussian Process                                                 & \begin{tabular}[c]{@{}c@{}}Monte Carlo  Dropout\end{tabular}                                & Ensemble Method                     \\ \hline
\begin{tabular}[c]{@{}c@{}}UQ Type\end{tabular}             & Epistemic (CI)                                                       & Epistemic (CI)                                                                                   & Epistemic (CI)                    \\ \hline
\begin{tabular}[c]{@{}c@{}}Visualized \\ Bounds\end{tabular}            & $[\pm1\sigma, \pm2\sigma, \pm3\sigma]$                                    & $[\pm1\sigma, \pm2\sigma, \pm3\sigma]$                                                                & $[\pm1\sigma, \pm2\sigma, \pm3\sigma]$ \\ \hline
\begin{tabular}[c]{@{}c@{}}Coverage Rate \\ for Each Bound\end{tabular} & [10.2\%, 18.2\%, 29.1\%]                                                     & [10\%, 20.8\%, 36.4\%]                                                                                 & [11.1\%, 23.3\%, 36.3\%]            \\ \hline
RMSE/$R^2$                                                              & 113.1452/0.2429                                                         & 56.2704/0.8127                                                                                     & 53.9461/0.8279               \\ \hline
Others                                                                  & Assume the noise is homeostatic                                                      & \begin{tabular}[c]{@{}c@{}}Dropout rate = 0.2;\\ MC sample=1e5\end{tabular}                  & Num. of NNs: 10                \\ \hline \hline                                                                 & \includegraphics[width=0.27\linewidth]{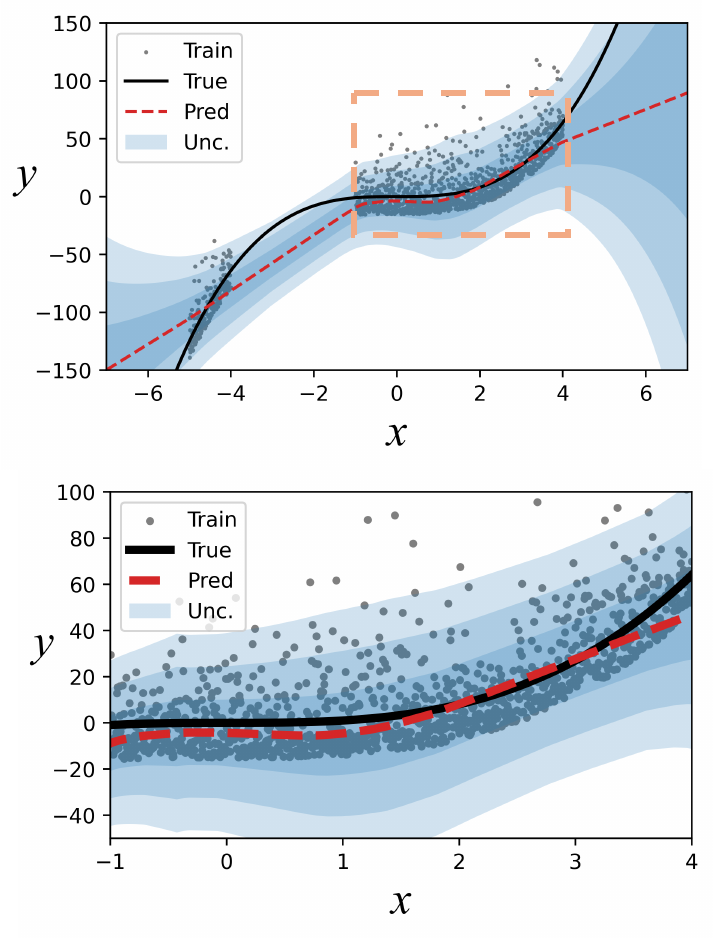}                                                          & \includegraphics[width=0.27\linewidth]{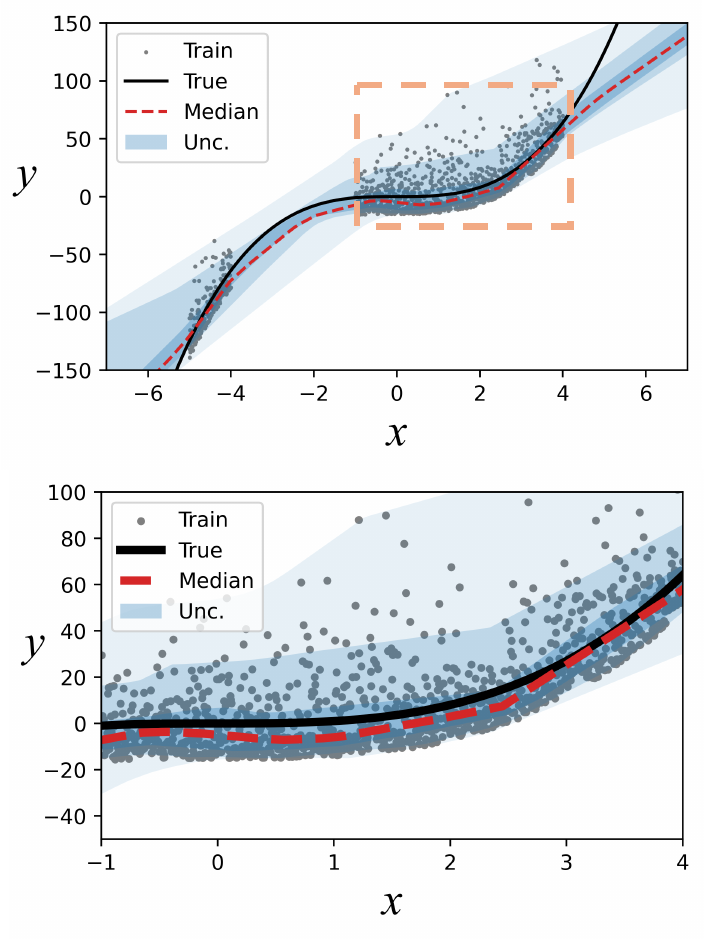}                                                                                      &                               \\ 
Method                                                                  & \begin{tabular}[c]{@{}c@{}}Evidential Regression\end{tabular} & \begin{tabular}[c]{@{}c@{}}Quantile Regression\end{tabular}                               &                               \\ \cline{1-3}
\begin{tabular}[c]{@{}c@{}}UQ Type\end{tabular}             & \begin{tabular}[c]{@{}c@{}}Epistemic + Aleatoric (PI)\end{tabular} & Aleatoric (PI)                                                                                    &                               \\ \cline{1-3}
\begin{tabular}[c]{@{}c@{}}Visualized \\ Bounds\end{tabular}            & $[\pm1\sigma, \pm2\sigma, \pm3\sigma]$                                    & \begin{tabular}[c]{@{}c@{}}$q$ = {[}(0.25, 0.75), (0.05, 0.95), \\(0.001, 0.999){]}\end{tabular} &                               \\ \cline{1-3}
\begin{tabular}[c]{@{}c@{}}Coverage Rate \\ for Each Bound\end{tabular} & {[}40.2\%, 68.4\%, 85.8\%{]}                                      & [32.7\%, 59.5\%, 73.2\%]                                                                                &                               \\ \cline{1-3}
RMSE/$R^2$                                                              & 73.7554/0.6783                                                   & 53.8638/0.8218                                                                                     &                               \\ \cline{1-3}
Others                                                                  & $\lambda=1e-4$                                                   & \begin{tabular}[c]{@{}c@{}}Each quantile is predicted \\ by an individual NN\end{tabular}                                                                                   &                               \\ \cline{1-3} 
\end{tabular}
\end{table*}

As can be seen from the results in Table \ref{table:UQ_benchmark}, because GP, MC dropout, ensemble method, and evidential regression all assumes the uncertainty distribution to be normal, only quantile regression can effectively capture the skewed distribution, while others failed to provide accurate uncertainty bounds. In fact, since GP, MC dropout, and ensemble method are quantifying epistemic (model) uncertainty, and are predicting confidence intervals (CI) rather than predictive intervals (PI), it is reasonable that these methods failed to capture the aleatoric (data) uncertainty, which is more critical in our application. Evidential regression, by learning the uncertainty of the estimated standard deviation, is capable of learning both epistemic and aleatoric uncertainty. However, similarly, it suffers from assuming the distribution to be normal, and provides obvious over-conservative CIs. In contrast, by using quantile regression, it can flexibly learn median response and bounds effectively regardless of the uncertainty distribution. The prediction accuracy of quantile regression also outperform other methods.

These results supports the need of quantile regression as an computationally efficient, flexible, accurate UQ methods for applications in robust MPC for nonlinear systems.

\subsection{Implementation of Quantile Regression on TiDE}

The quantile loss for time-series data is an extension of the standard quantile loss. For time series data with $N$ steps ahead to be predicted, the total quantile loss is often calculated as the sum over all time steps and quantile levels:
\begin{equation}
    L_{Q} = \frac{1}{N\cdot l}\sum_{j=1}^{l}\sum_{t=1}^{N}L_{q,j}(y_t,\hat{y}_t),
\end{equation}
where $l$ is the levels of the assigned quantiles. 

Implementing quantile loss in TiDE involves increasing the output dimensions to accommodate state predictions for multiple quantile levels. For a prediction setup with batch size $B$, horizon length $N$, number of responses $D$, and $l$ quantile levels for each response, the output tensor from TiDE will have dimensions $[B, N, D, l]$. This expanded output structure facilitates the direct estimation of uncertainty bounds by providing multiple quantile estimates for each prediction. This design allows TiDE to predict the entire output tensor, including all quantile values for each response, in a single forward pass. The TiDE prediction model with the quantile output can be represented by:
\begin{align}
    \hat{\mathbf{x}}_{k+1:k+N} & = [\bar{\hat{\mathbf{x}}}_{k+1:k+N},\tilde{\hat{\mathbf{x}}}_{k+1:k+N}, \underline{\hat{\mathbf{x}}}_{k+1:k+N}]^T \notag  \\ & = \text{TiDE}(\mathbf{x}_{k-w+1:k},\mathbf{d}_{k-w:k-1}, \mathbf{u}_{k-w:k-1}, \mathbf{d}_{k:k+N-1},\mathbf{u}_{k:k+N-1} | \boldsymbol{\phi}). 
    \label{eq:TiDE_with_quantile}
\end{align}

In this context, $\bar{\hat{\mathbf{x}}}_{k:k+N-1},\tilde{\hat{\mathbf{x}}}_{k:k+N-1}, \underline{\hat{\mathbf{x}}}_{k:k+N-1}$ represent the upper quantile, median, and lower quantile of the predicted states, respectively. To simplify the notation, we utilize superscripts $p$, $f$, and $p:f$ to denote the past, the future, and the past and future covariates/targets, respectively, at time $k$. The equation of the TiDE model becomes:
\begin{equation}
[\bar{\hat{\mathbf{x}}}^f_{k+1},\tilde{\hat{\mathbf{x}}}^f_{k+1}, \underline{\hat{\mathbf{x}}}^f_{k+1}]^T = \text{TiDE}(\mathbf{x}_{k}^p, \mathbf{u}_{k}^p,\mathbf{u}_{k}^f, \mathbf{d}_{k}^p, \mathbf{d}_{k}^f|\boldsymbol{\phi}).
\label{eq:TiDE_with_quantile_simplified}
\end{equation}
By employing this one-shot approach, TiDE substantially reduces computation time through efficient parallelization. 

\section{Simultaneous Multi-Step Robust MPC for Digital Twin} 


The purpose of this work is to integrate the simultaneous multi-step ahead predictive quantile with MPC, demonstrating a learning-based robust MPC that performs decision-making under uncertainty for Digital Twins. In this section, we first describe the development of the virtual system via TiDE. Then, we focus on the robust MPC formulation incoperating multi-step quantile prediction, and introduce the optimization tools that can benefit real-time solving. A complete walk-through of the process will be detailed in Section \ref{sec:example}.

\subsection{Building the Virtual System via TiDE: Data Generation and Model Training}

TiDE serves not only to learn the dynamics of the physical system but also to capture the distribution of system behaviors under the influence of uncertainty. To achieve this, the training data collected from the physical system must accurately represent its behavior under operational disturbances. This ensures that the model can generalize effectively and provide reliable predictions across a range of operating conditions.

In many existing learning-based robust MPC approaches (when the system model is unknown), the disturbances affecting the system dynamics are assumed to be known beforehand, and training data are typically generated through virtual experiments under predefined disturbance conditions. However, in practical engineering scenarios where the distribution of uncertainties is unknown, the collected data inherently includes noise and can be directly utilized as training data for TiDE, enabling it to adapt to real-world conditions. This data-driven approach allows TiDE to model both the nominal system behavior and the variability introduced by stochastic disturbances.

At the model training stage, TiDE directly learns the user-assigned quantile levels from the noisy training data. In this setting, there is no assumption made regarding the boundedness of the response under uncertainty. Consequently, even trained under noisy data, TiDE can provide smooth predictions for the quantile levels and the median, efficiently quantifying the data (aleatoric) uncertainties. This allows the model to balance prediction accuracy with uncertainty estimation, making it suitable for robust decision-making in dynamic systems. This is because the crucial features are extracted and mapped into the dense encoder, which also serves as a noise filter to eliminate the impact of noise and disturbances while preserving the important information. When training the TiDE model using \texttt{PyTorch}, we use \texttt{Adam} optimizer as default, and add a regularization term to increase the generality of the model. We will exhibit specific details in the examples.

\subsection{Real-Time Decision-Making via MPC}


\subsubsection{Uncertainty-Aware MPC Formulation}
In robust MPC formulation in Eq.~(\ref{eq:RMPC_constraints}), the constraints are formulated as hard constraints to account for disturbances that are assumed to be bounded within a predefined set. The tube around the nominal trajectory is specified to ensure that all possible realizations of the bounded disturbances remain within this tube, as shown in Fig.~\ref{fig:MPC_illustration}(c). However, for those disturbances which are not bounded (e.g., Gaussian noises), it is nearly impossible to guarantee the satisfaction of hard constraints. In this case, the constraints are relaxed into probabilistic (chance) constraints, ensuring that they are satisfied with a specified probability~\cite{heirung2018stochastic}. Therefore, the tubes are derived probabilistically based on the distribution of the disturbances~\cite{cannon2009probabilistic}. This approach acknowledges that disturbances may not have strict bounds but instead follow a known or estimated probability distribution. When a confidence level $\alpha$ (e.g., $\alpha=0.95$) is specified, the tube bounds can be explicitly calculated to encapsulate 95\% of the disturbance realizations. This probabilistic bounding allows for the construction of stochastic tubes that balance conservatism and feasibility, providing a probabilistic guarantee of constraint satisfaction. Importantly, the probabilistic nature of the tube makes stochastic MPC less conservative than robust MPC while still accounting for uncertainty effectively.

Let us assume that the constraints are only enforced on states and control input. The uncertainty-aware MPC with the single-step nonlinear dynamics and probabilistic constraints can be formulated as:
\begin{subequations}
\begin{align}
    \min_{\mathbf{u}} ~&\mathbb{E}{\mathbf{w}}[J(\mathbf{u},\mathbf{x},\mathbf{r},\mathbf{w})], \label{eq:RMPC_prob_objective} \\
    s.t. \quad & \mathbf{x}_{k+i+1}={F}_w(\mathbf{x}_{k+i},\mathbf{u}_{k+i},\mathbf{w}_{k+i}),~\forall i\in\mathbb{N}_{[0,N-1]},\label{eq:RMPC_prob_constraints_b}\\
       & \text{Pr}\left({x}^{(j)}_{k+i}\in\mathbb{X}\right)\geq\alpha,~\forall i\in\mathbb{N}_{[0,N-1]},~\forall j\in\mathbb{N}_{[1,n_x]},\label{eq:RMPC_prob_constraints_d}\\
      &  \mathbf{u}_{k+i}\in\mathbb{U},~\forall i\in\mathbb{N}_{[0,N-1]},\label{eq:RMPC_prob_constraints_e}
    \end{align}
    \label{eq:RMPC_prob_constraints}
\end{subequations}
where $\mathbf{w}=[\mathbf{w}_k,...,\mathbf{w}_{k+p-1}]$ represents the sequence of the random variables for the disturbance vectors and $n_x$ denotes the dimension of the vector $\mathbf{x}$.


The optimization problem presented in Eq. (\ref{eq:RMPC_prob_constraints}) poses significant computational challenges due to its probabilistic constraints and the need to consider uncertainty propagation across the prediction horizon. While the formulation elegantly captures the uncertainty-aware nature of the control problem, its direct implementation is computationally intractable, particularly for real-time applications.

\subsubsection{Reformulation with Quantiles and Constraint Tightening}
To address this computational challenge, the uncertainty-aware MPC problem is reformulated using the quantile information and constraint tightening techniques with an ancillary controller design. The predictive model in Eq. (\ref{eq:TiDE_with_quantile_simplified}) provides information about the upper and lower bounds of future states, which can be directly utilized to ensure constraint satisfaction under uncertainty. The problem from Eq.~(\ref{eq:RMPC_prob_constraints}) can be reformulated as:
\begin{subequations}
\begin{align}
    \min_{\mathbf{v}} ~&J(\mathbf{v},\hat{\mathbf{x}}, \mathbf{r}), \label{eq:RMPC_prob_objective} \\
    s.t. \quad & [\bar{\hat{\mathbf{x}}}^f_{k},\tilde{\hat{\mathbf{x}}}^f_{k}, \underline{\hat{\mathbf{x}}}^f_{k}]^T = \text{TiDE}(\mathbf{x}_{k}^p, \mathbf{u}_{k}^p,\mathbf{u}_{k}^f, \mathbf{d}_{k}^p, \mathbf{d}_{k}^f|\boldsymbol{\phi})\label{eq:RMPC_prob_constraints_tightened_b}\\
       & \bar{\hat{x}}_{j,k+i}\leq x_{j,\text{ub}},~\forall i\in\mathbb{N}_{[0,N-1]},~\forall j\in\mathbb{N}_{[1,n_x]},\label{eq:RMPC_prob_constraints_tightened_c}\\
       & \underline{\hat{x}}_{j,k+i}\geq x_{j,\text{lb}},~\forall i\in\mathbb{N}_{[0,N-1]},~\forall j\in\mathbb{N}_{[1,n_x]},\label{eq:RMPC_prob_constraints_tightened_d}\\
      &  \mathbf{v}_{k+i}\in\mathbb{U}\ominus\mathbf{K}\mathbb{Z}_{k+i},~\forall i\in\mathbb{N}_{[0,N-1]},\label{eq:RMPC_prob_constraints_tightened_e}\\
      &\mathbf{u}_{k}=\mathbf{v}_{k}+\mathbf{K}\mathbf{e}_{k},\label{eq:RMPC_prob_constraints_tightened_f}
    \end{align}
    \label{eq:RMPC_prob_constraints_tightened}
\end{subequations} 
where the subscripts `ub' and `lb' denote the upper and lower bounds, respectively, the subscript $j$ represents the index of the state vector $\mathbf{x}$, $\mathbf{v}=\left[\mathbf{v}_k,\mathbf{v}_{k+1},...,\mathbf{v}_{k+N-1}\right]$ denotes the sequence of the nominal control inputs (which are the decision variables), $\mathbf{e}_{k}=\mathbf{x}_{k}-\tilde{\hat{\mathbf{x}}}_k$ represents the deviation between the actual and predicted states, and $\mathbb{Z}_{k+i}$ is the set of the quantile bound at time $k+i$. 

Theoretically, the optimal value of $\mathbf{K}$ can be determined using the LQR for linear tube-based robust MPC \cite{rawlings2017model, tsai2023robust}. This value plays a crucial role in stabilizing the system and preventing the estimation error bound from diverging during multi-step-ahead prediction. It also influences the level of conservativeness in estimating error bounds when propagating noise through recursive rollout. However, for nonlinear (and even unknown) systems, LQR cannot be used to obtain $\mathbf{K}$, and it must be fine-tuned or optimized using black-box methods such as Bayesian optimization \cite{chen2024latent}. It’s important to note that optimizing $\mathbf{K}$ is beyond the scope of this work. In contrast, TiDE directly captures the actual distribution from the open-loop data, unlike tube-based robust MPC. Therefore, the value of $\mathbf{K}$ does not affect the conservativeness of the predicted error bounds. In online solving, the value of $\mathbf{K}$ tightens the design space of $\mathbf{u}_k$ to account for system disturbances. This is demonstrated in the example provided, which can also be derived from the Riccati equation \cite{tsai2023robust}. A more rigorous implementation would involve continuous linear approximations \cite{schwedersky2022nonlinear} on the TiDE model and deriving $\mathbf{K}$. However, this is also beyond the scope of our work.

The state constraints from Eq. (\ref{eq:RMPC_prob_constraints_tightened_c}) and Eq. (\ref{eq:RMPC_prob_constraints_tightened_d}) are managed through the quantile information derived from the predictive model. By utilizing these bounds, we can guarantee that the system states remain within their feasible region with the specified probability level $\alpha$. This approach effectively transforms the probabilistic state constraints into deterministic bounds based on the predicted quantiles of the state distribution.

For the control input constraints in Eq. (\ref{eq:RMPC_prob_constraints_tightened_e}), a more careful strategy is necessary to ensure that the actual applied control actions remain within the physical limitations of the controller or actuator. The control input constraints are tightened~\cite{lorenzen2016constraint,gonzalez2020comparative,gao2014tube} to accommodate the additional control effort that may be required by the ancillary controller in Eq. (\ref{eq:RMPC_prob_constraints_tightened_f}). The ancillary controller is used for rejecting the real-time disturbance while maintaining the satisfaction of the original constraints~\cite{lopez2019dynamic}. In this study, a linear representation is chosen for computational efficiency. The constraint-tightening technique creates a safety margin that can avoid the actual state $\mathbf{x}_k$ and the total control input $\mathbf{u}_k$ violating the original constraints $\mathbb{X}$ and $\mathbb{U}$, respectively.

\subsection{Optimization Setup}

Although one motivation of simultaneous multi-step MPC is to accelerate the solving process of MPC by parallelizing the state prediction in one-shot, in this work, we implement other techniques to further accelerate the optimization process, making the MPC solvable in actionable time. Here we detail the methods and algorithm used to accelerate the solving process of MPC. 

\subsubsection{Gradient-Based Optimization with Automatic-Differentiation}
\begin{figure}[h!]
    \centering
\includegraphics[width=0.5\linewidth]{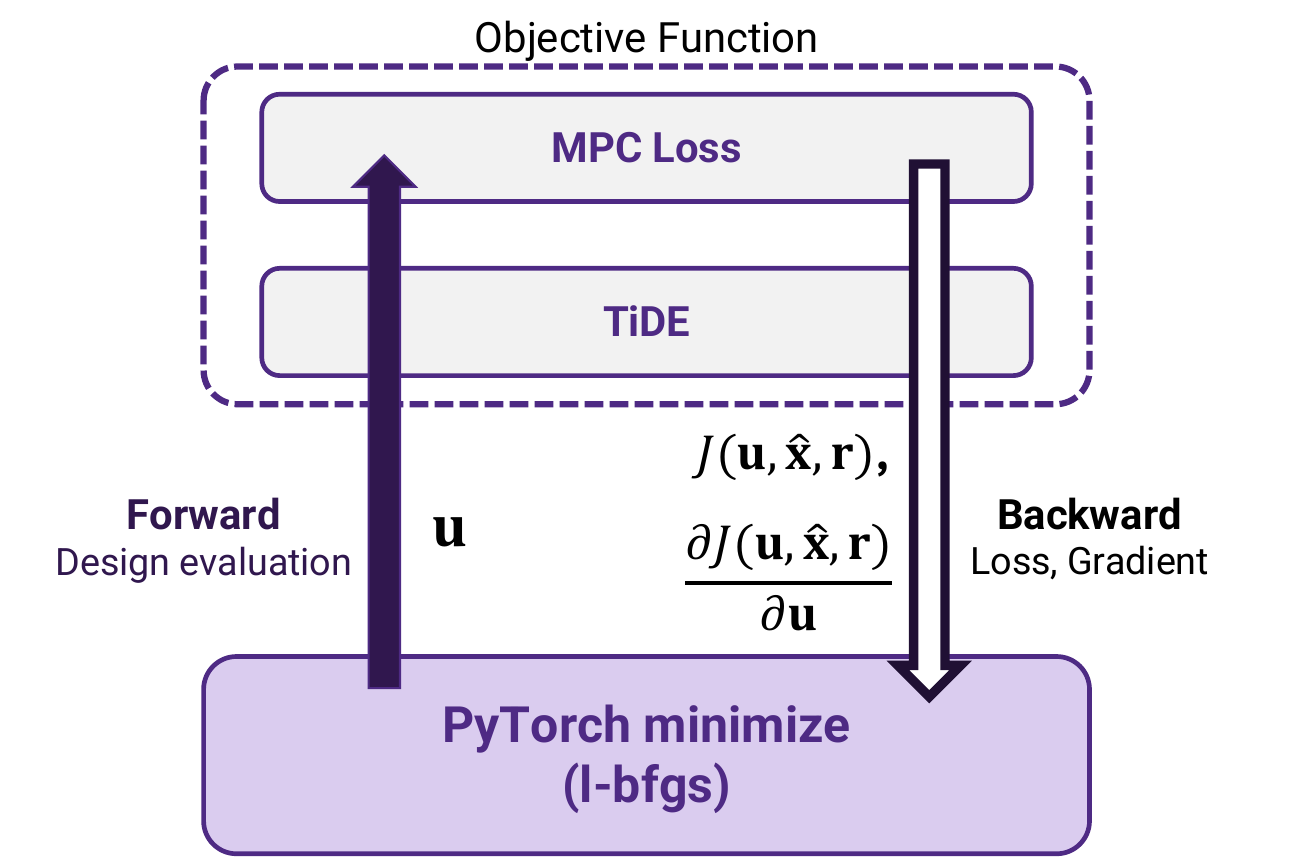}
    \caption{Illustration of gradient-based optimization using auto-differentiation.}
    \label{fig:autograd}
\end{figure}

One way to accelerate the solving process of MPC using a numerical optimization solver is to apply gradient-based optimization with automatic differentiation \cite{jung2023model}. The key idea is to acquire analytical evaluation of the first-order derivative of the loss function with respect to the design variables (control input $\mathbf{u}$), and use the gradient information to perform gradient-based optimization. Since the evaluation of the MPC loss $J(\mathbf{u},\hat{\mathbf{x}}, \mathbf{r})$ as well as TiDE are both computed using \texttt{PyTorch} \cite{paszke2019pytorch}, the gradient of MPC loss $\partial J(\mathbf{u},\hat{\mathbf{x}}, \mathbf{r})/\partial \mathbf{u}$ can be obtained analytically using backpropagation instead of numerical approximations such as finite difference, as shown in Fig. \ref{fig:autograd}. Lastly, in this work, we choose \texttt{l-bfgs} \cite{nocedal1999numerical} with a \texttt{Pytorch} wrapper developed by \cite{Feinman2021} as MPC's numerical optimizer. The \texttt{l-bfgs} is a light memory-used algorithm that approximates the Hessian (second-order derivative) using the first-order derivative of the loss function. Since the gradient information can be obtained automatically, the evaluation of Hessian can also be done in only one function evaluation. As a result, the integration of \texttt{l-bfgs} and \texttt{Pytorch} enables efficient gradient-based optimization by utilizing cheap but accurate gradient evaluation.

\subsubsection{Penalty Method: Augmented Lagrangian Method}

However, even when the gradient of the objective function can be automatically computed, evaluating the optimality conditions in \texttt{l-bfgs} when constraints are enforced still requires additional numerical approximations of second-derivative terms, resulting in a significant increase in function evaluation time and computational speed. Therefore, we employ the penalty method to transform the constraint optimization problem into an unconstrained optimization problem. By directly incorporating the penalty terms into the objective function, we convert the hard constraints into soft constraints, thereby bypassing the evaluation the optimality conditions associated with constraint optimization. 

In particular, we use the augmented Lagrangian method \cite{nocedal1999numerical} for the penalty method. Assume the objective function for the constrained MPC/robust MPC is $J(\mathbf{u},\hat{\mathbf{x}}, \mathbf{r})$ with constraint $c_i(\mathbf{u},\mathbf{x}), i \in \mathcal{E}$ generalized for both equality and inequality constraints, where $\mathcal{E}$ is the number of constraints. The augmented Lagrangian method then solves the following unconstrained optimization problem by adding the constraints as a penalty:
\begin{align}
    \min_{\mathbf{u}} \Phi_{s}(\mathbf{u},\hat{\mathbf{x}}, \mathbf{r}) = J(\mathbf{u},\hat{\mathbf{x}}, \mathbf{r}) & + \frac{\mu_i}{2}\sum_{i\in\mathcal{E}}[\text{ReLU}(c_i(\mathbf{u},\hat{\mathbf{x}}))]^2 +\sum_{i\in\mathcal{E}}\lambda_i \text{ReLU}(c_i(\mathbf{u},\hat{\mathbf{x}})), \label{eq:augmented_lag_obj}
\end{align}
where $s$ indicates the $s$th iteration when solving the optimization problem. Both $\mu_i$ and $\lambda_i$ are the penalty 
 parameter and the estimated Lagrange multiplier corresponding to the $i$th constraints and follow the updating rules:
\begin{equation}
    \mu_i \leftarrow \alpha\mu_i,
\end{equation}
\begin{equation}
    \lambda_i \leftarrow \lambda_i + \text{ReLU}(c_i(\mathbf{u}_s,\hat{\mathbf{x}})),
\end{equation}
where $\alpha>1$ is the increasing rate of $\mu_i$, and $\mathbf{u}_s$ is the solution for solving unconstrained optimization (Eq. (\ref{eq:augmented_lag_obj})) at iteration $s$. In this work, we select $\lambda_0=10, \mu_0 = 1$, and $\alpha=3$ from trial and errors that balance constraint satisfaction and convergence rate. In the next iteration, the solver will resolve the problem using $\mathbf{u}_s$ as the initial guess for warm start. Here, we apply the $\text{ReLU}()$ function in Eq. (\ref{eq:augmented_lag_obj}) because it is a continuous and differentiable function that only penalizes $\Phi_{s}$ when the constraints $c_i$ are violated, and enabling the smooth computation of the gradient of $\Phi_s$. 

While the augmented Lagrangian method introduces some deviation from strict KKT conditions to improve numerical stability and feasibility, it still provides robust handling of both equality and inequality constraints. Further, it avoids the ill-conditioning issues of pure penalty methods by using Lagrange multipliers, reducing sensitivity to the penalty parameter. Lastly, it converges more efficiently to feasible solutions, even for problems with non-linear constraints \cite{nocedal1999numerical}. 

Lastly, the warm start is used to provide a potential starting point near the optimal solution, i.e., the optimal solution from the previous step is used as the initial guess for the current step. Also, if warm starting MPC is unable to identify a feasible solution, the MPC will terminate the optimization and use the pre-defined control input to achieve feasibility of online operation.

\section{Illustrative Example}
\label{sec:example}
We first verify the proposed method using a linear invariant system so that the result can be compared with the tube-based method, which is one of the most widely adopted robust MPC methods. This section provides a complete walk-through by introducing the physical system, developing the virtual system, building the virtual-to-physical connection via robust MPC, and result validation and comparison.

\subsection{Physical System}

A discrete linear system is selected for demonstration. The system with exogenous noise on input $\epsilon_k\sim\mathcal{N}(0,0.1^2)$ is formulated as follows:

\begin{align}
     \mathbf{x}_{k+1} & = \begin{bmatrix}
0.3 & 0.1 \\
0.1 & 0.2 \\
\end{bmatrix} \mathbf{x}_k + \begin{bmatrix}
0.5 \\ 1
\end{bmatrix} \left( u_k + \epsilon_k \right)  = F_w(\mathbf{x}_k, u_k, \mathbf{w}_k),
\label{eq:toy_problem_sys}
\end{align}
where the disturbance vector is the multiplication of matrix $\mathbf{B}$ and the noise vector $\boldsymbol{\epsilon}_k$, i.e., $\mathbf{w}_k=\mathbf{B}\boldsymbol{\epsilon}_k$, which also follows a Gaussian.

\subsection{Virtual System}
\subsubsection{Data Generation and Model Training}
The development of the virtual system in our work follows the steps of Stage 1 in Fig. \ref{fig:overall_flowchart}. To generate state trajectories for system identification using TiDE, a sequence of input $\mathbf{D}_{u} = \{u_0,\hdots,u_{n-1}\}$, where $\mathbf{u} \in [-5,5]$ is uniformly sampled with a size of $n = 422,000$. By setting the initial state $\mathbf{x}_0 = [0,0]^T$, the trajectory of $\mathbf{D}_x = \{\mathbf{x}_{1},\hdots,\mathbf{x}_{n}\}$ can be simulated by using $\mathbf{u}$ as the input. Furthermore, both $\mathbf{D}_u$ and $\mathbf{D}_x$ are divided into fractions using the moving window approach, with each fraction having a length of $w+N$. In this case, the window size is $w=10$, and the horizon length is $N=10$. By denoting the $l$th fraction as $\mathbf{D}^l_u = \{u_{l},\hdots,u_{l+w+N-1}\}$ and $\mathbf{D}^l_x = \{\mathbf{x}_{l},\hdots,\mathbf{x}_{l+w+N-1}\}$, respectively, we can assign $\mathbf{x}_{p} = [\mathbf{x}_{l+1},\hdots,\mathbf{x}_{l+w}]$ as the past target, $\mathbf{x}_{f}=[\mathbf{x}_{l+w+1},\hdots,\mathbf{x}_{l+w+N}]$ as the future target, with $\mathbf{u}_{p} = [u_{l},\hdots,u_{l+w-1}]$ and $\mathbf{u}_{f} = [u_{l+w},\hdots,u_{l+w+N-1}]$ as the past and future covariates (inputs), respectively. Then, $\mathbf{D}^l_x$ and $\mathbf{D}^l_u$ is further split into training, validation, and test set with a 8:1:1 ratio. Following Eq. (\ref{eq:TiDE_general}), the TiDE model for identifying this system can be trained via supervised learning using quantile loss, which is formulated as:
\begin{subequations}
\begin{align}
    \min_{\boldsymbol{\phi}} \quad & L_Q(\mathbf{x}^f, \hat{\mathbf{x}}^f) \\
    s.t. \quad & \hat{\mathbf{x}}^f = \text{TiDE}(\mathbf{x}^p, \mathbf{u}^p,\mathbf{u}^f | \boldsymbol{\phi}).
\end{align}
\end{subequations}
The detail of the training and model set up is described in Table \ref{tab:tide_training}. The training and validation loss is shown in Fig. \ref{fig:TiDE_evaluation_toy}(a).

\begin{table*}[t]
\centering
\caption{Hyperparamters of training TiDE model}
\label{tab:tide_training}
\small
\begin{tabular}{ccccccc}
\hline \hline
\multicolumn{7}{c}{Details for TiDE model setup}                                                                                                                                                                                                                         \\ \hline
\multicolumn{1}{c|}{\# encoder layers} & \multicolumn{1}{c|}{\# decoder\_layers} & \multicolumn{1}{c|}{decoder output dim.} & \multicolumn{1}{c|}{hidden size} & \multicolumn{1}{c|}{decoder hidden size} & \multicolumn{1}{c|}{dropout rate} & layer normalization \\ \hline
\multicolumn{1}{c|}{1}                 & \multicolumn{1}{c|}{1}                  & \multicolumn{1}{c|}{16}                  & \multicolumn{1}{c|}{128}         & \multicolumn{1}{c|}{32}                           & \multicolumn{1}{c|}{0.2}          & True                \\  \hline \hline
\multicolumn{7}{c}{Details for TiDE model training}                                                                                                                                                                                                                            \\ \hline
\multicolumn{1}{c|}{learning rate}     & \multicolumn{1}{c|}{regularization}     & \multicolumn{1}{c|}{step\_size}          & \multicolumn{1}{c|}{rate decay}  & \multicolumn{1}{c|}{\# epoch}                     & \multicolumn{1}{c|}{batch size}   & shuffle data        \\ \hline
\multicolumn{1}{c|}{0.001}             & \multicolumn{1}{c|}{0.002}              & \multicolumn{1}{c|}{10}                   & \multicolumn{1}{c|}{0.95}        & \multicolumn{1}{c|}{1500}                          & \multicolumn{1}{c|}{64}          & True                \\ \hline \hline
\end{tabular}
\end{table*}

\subsubsection{Model Evaluation}
\begin{figure}[h!]
    \centering
    \includegraphics[width=0.6\linewidth]{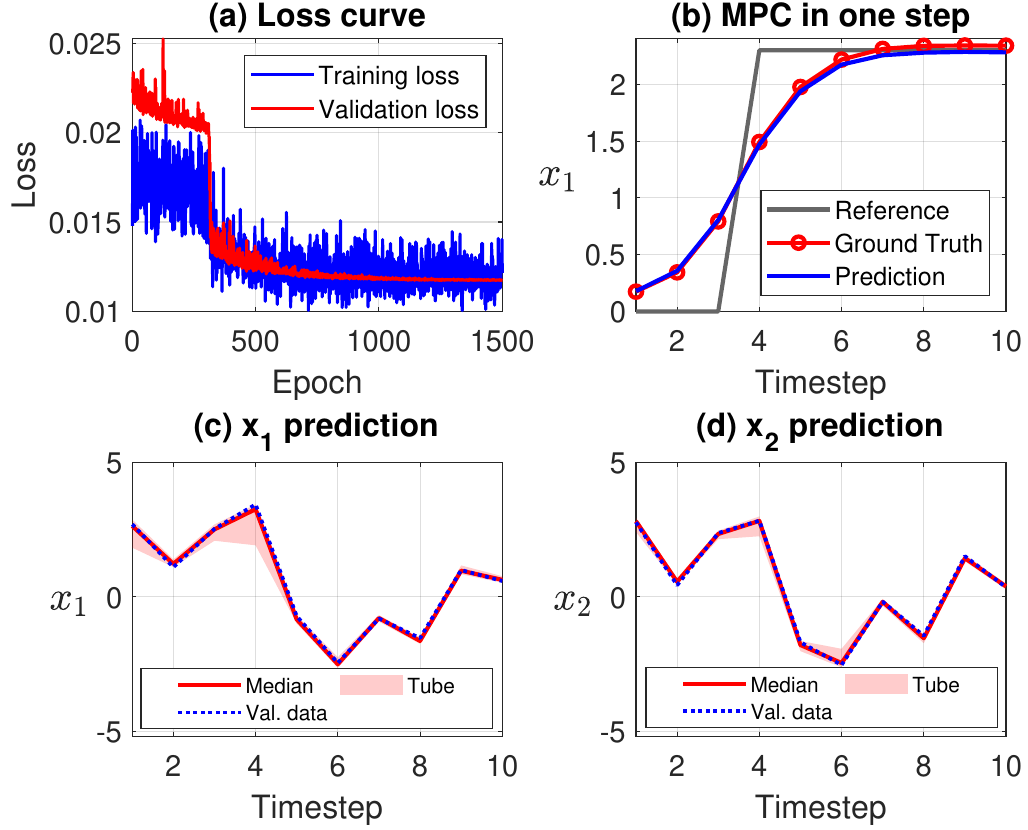}
    \caption{Evaluation of TiDE. (a) The training and validation loss of TiDE. (b) Comparison of the state prediction and the ground truth in a single MPC step. (c) One-shot prediction of the median and quantiles of $x_1$. (d) One-shot prediction of the median and quantiles of $x_2$. }
    \label{fig:TiDE_evaluation_toy}
\end{figure}

We first evaluate the accuracy of TiDE using the test set. Quantitatively, the mean absolute percentage error (MAPE) and the relative residual mean square error (RRMSE) for the predicted $x_1$ and $x_2$ achieves $[5.96\%, 5.05\%]$ and $[0.0419, 0.0414]$, respectively, showing the high predictive capability of TiDE. Qualitatively, as shown in Fig. \ref{fig:TiDE_evaluation_toy}(c) and \ref{fig:TiDE_evaluation_toy}(d), the tube, (i.e., the error interval bounded by the 0.95 and 0.05 quantiles), as well as the median of the predicted of $x_1$ and $x_2$ by TiDE is compared with the validation data (with noise injected) using a randomly selected fraction in the test set. As can be seen from the figure, the predicted median matches the validation data well with slight deviation, and the validation data is mostly bounded by the upper and lower quantiles. This shows the capability of TiDE to capture the dynamics of the system, and to quantify the uncertainty of the prediction even when the training data is noisy.

In Figure \ref{fig:TiDE_evaluation_toy}(b), we compare the predicted median of $x_1$ $\tilde{\hat{x}}_{1,k+1:k+N}$ and the ground truth (obtained from the system model without injected noise) using the optimal control input $\mathbf{u}_k^*$ obtained by solving Eq. (\ref{eq:MPC_toy}) with a step function as the reference. Here, we demonstrate that the optimizer successfully solves the MPC problem that minimizes the reference tracking error using the multi-step ahead prediction by TiDE. By examining the results, we observe that the predicted response shares a similar trend with the true response. This test validates the optimization capability of TiDE as the multi-step ahead predictor in MPC. 

\begin{figure}[t]
    \centering
    \includegraphics[width=1\linewidth]{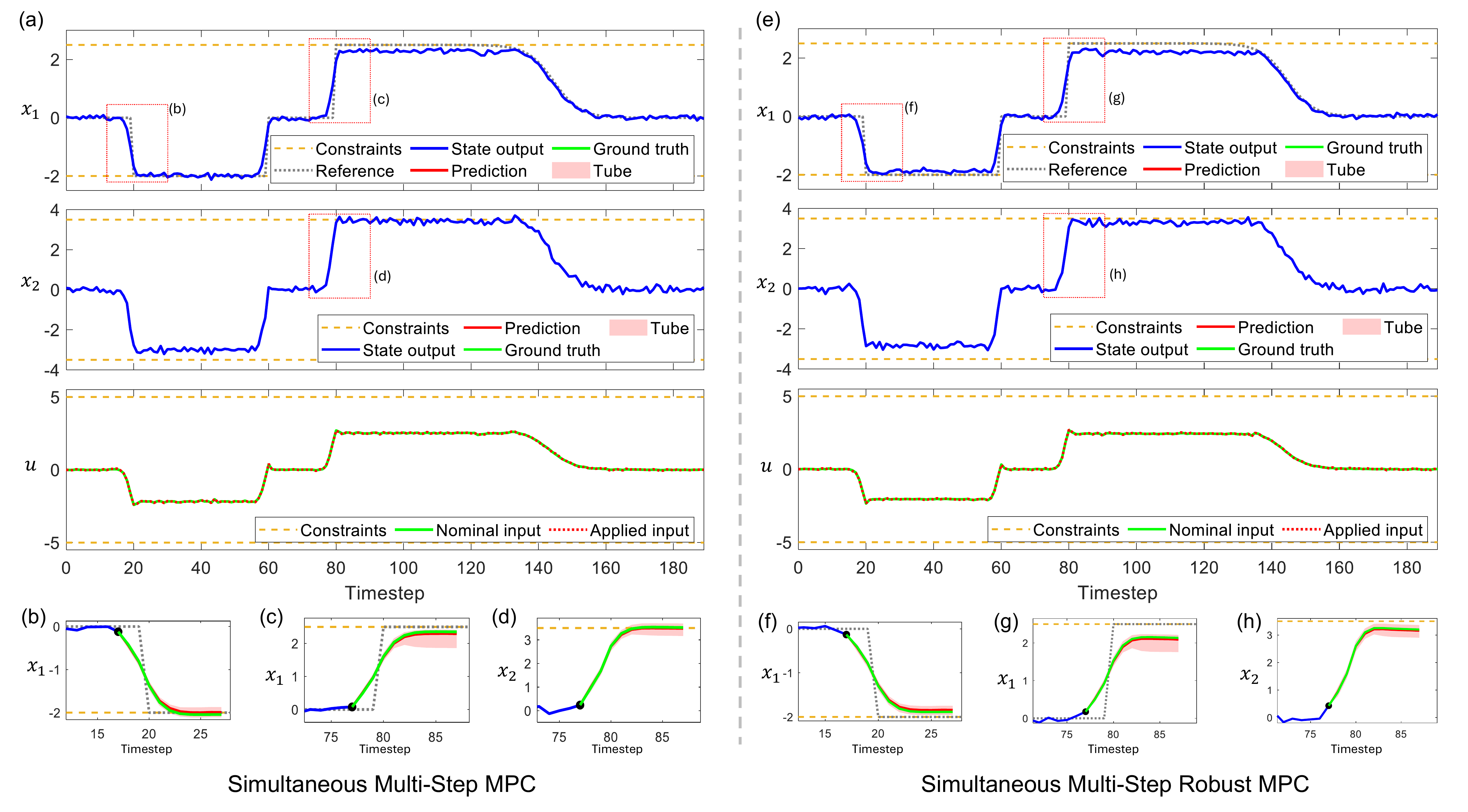}
    \caption{Comparison of simultaneous multi-step MPC with and without robust consideration. (a) Trajectories of the states and input under multi-step (nominal) MPC. (b), (c), and (d) show the selected highlights of $x_1$ and $x_2$ from (a) where the constraints are active. (e) Trajectories of the states and input under multi-step robust MPC. (f), (g), and (h) show the selected highlights of $x_1$ and $x_2$ from (e) where the constraints are active.}
    \label{fig:toy_rmpc_compare_single}
\end{figure}

\subsection{Virtual-to-Physical Integration}
 The bi-directional interaction between the physical and virtual system is constructed with robust MPC, where the current state of the physical system is feedback to the virtual system as its prediction input. Further, an optimal control input is solved by an online robust MPC, using the predicted quantiles as the safety buffer to explicitly handle constraints under uncertainty. Then, the optimal control input is applied to the physical system.
 
 The goal of this example is to use robust MPC to perform a reference tracking task on $x_1$, while maintaining $x_1$, $x_2$, and $u$ within the feasible regions subjected to unbounded disturbance. The multi-step robust MPC using the quantile prediction from TiDE can be formulated as:
\begin{subequations}
\label{eq:MPC_toy}
\begin{align}
    \min_{\mathbf{v}} \quad & J(\mathbf{v},\tilde{\hat{\mathbf{x}}},\mathbf{r}) = \sum_{i=0}^{N-1} ||\tilde{\hat{x}}_{1,k+i+1} - r_{k+i+1}||_{\mathbf{Q}}^2+||v_{k+i}||_{R}^2\\
    s.t. \quad & \hat{\mathbf{x}}_{k+1}^f=[\bar{\hat{\mathbf{x}}}_{k+1}^{f},\tilde{\hat{\mathbf{x}}}_{k+1}^{f}, \underline{\hat{\mathbf{x}}}_{k+1}^{f}]^T = \text{TiDE}(\mathbf{x}_{k}^p, \mathbf{u}_{k}^p,\mathbf{v}),\\
    & \mathbf{v} = [v_k,\hdots,v_{k+N-1}], \\
    & \text{Pr}\left(\hat{x}_{1,k+i} \geq -2 \right) \geq 0.95, \forall i \in \mathbb{N}_{[1,N]}, \label{eq:pr_g_1_toy_ub}\\
    & \text{Pr}\left(\hat{x}_{1,k+i} \leq 2.5 \right) \geq 0.95, \forall i\in\mathbb{N}_{[1,N]}, \label{eq:pr_g_1_toy_lb}\\
    & \text{Pr}\left(\hat{x}_{2,k+i} \geq -3.5 \right) \geq 0.95, \forall i\in\mathbb{N}_{[1,N]}, \label{eq:pr_g_2_toy_ub}\\
    & \text{Pr}\left(\hat{x}_{2,k+i} \leq 3.5 \right) \geq 0.95, \forall i\in\mathbb{N}_{[1,N]}, \label{eq:pr_g_2_toy_lb}\\
    & v_{k+i} \in \mathbb{U}\ominus \mathbf{K}\mathbb{Z}_{k+i}, \forall i\in\mathbb{N}_{[0,N-1]}, \\
    & u_k = \pi(v_k) = v_k + \mathbf{K}\mathbf{e}_k, \\
    & \mathbf{u}_{k}^p = [u_{k-w},...,u_{k-1}].
    \label{eq:u_g_toy}
\end{align}
\end{subequations}

Both $\mathbf{Q}$ and $\mathbf{R}$ are set to $\mathbf{I}$. The original input bound is $\mathbb{U}\in [-5,5]$, while it is dynamically tightened based on prediction error bound $\mathbb{Z}$. The linear regulator $\mathbf{K}$ = [-0.0621, -0.2027] is assigned in this case.  The Further, since the predicted upper and lower quantiles have already taken into account the probabilistic bounds on $\hat{x}_{1,k}$, the constraints in Equations (\ref{eq:pr_g_1_toy_ub})-(\ref{eq:pr_g_2_toy_lb}) can be rewritten as:
\begin{subequations} 
\begin{align}
    & \bar{\hat{\mathbf{x}}}_{1,k+i} \leq 2.5, \quad\underline{\hat{\mathbf{x}}}_{1,k+i} \geq -2,  \\
    & \bar{\hat{\mathbf{x}}}_{2,k+i} \leq 3.5, \quad \underline{\hat{\mathbf{x}}}_{2,k+i} \geq -3.5.
\end{align}
\end{subequations}

The performance of the proposed robust MPC method is compared with the multi-step MPC without uncertainty awareness (so-called nominal MPC) in Fig. \ref{fig:toy_rmpc_compare_single}. In this example, the reference trajectory is designed to overlap with the bound of $x_1$ and will violate the bound of $x_2$ during this reference tracking task, aiming to test the constraint handling capability of robust MPC at extreme scenarios. As can be seen from Fig. \ref{fig:toy_rmpc_compare_single}(a) where the nominal multi-step MPC is solved without considering the uncertainty of the prediction, the result yields significant constraint violation due to the disturbance. Since the gain of ancillary controller is small in this case, the applied input is almost identical to the nominal input. Fig. \ref{fig:toy_rmpc_compare_single}(b)-\ref{fig:toy_rmpc_compare_single}(d) compares the nominal TiDE prediction and the ground truth at different instances when the optimal control input sequence is solved. Here, the ground truth (green line) is verified by simulating Eq. (\ref{eq:toy_problem_sys}) without disturbance. As a result, even though the discrepancy between the nominal prediction and the ground truth is not significant, indicating the accuracy of the prediction, the optimal control input only provides ideal solutions that minimize the MPC loss but ignore the impact of uncertainty and does not provide safety buffer to accommodate disturbance.

In contrast, the proposed robust MPC method explicitly uses the learned quantile as the tube to provide an optimal control input that compromises the tracking performance in exchange for safety buffers to increase the chance of constraint satisfaction, as shown in Fig. \ref{fig:toy_rmpc_compare_single}(e). By taking a closer look at Fig. \ref{fig:toy_rmpc_compare_single}(f) and Fig. \ref{fig:toy_rmpc_compare_single}(g), we can see that the error bound (tube) is explicitly used when solving the constrained MPC, i.e., the robust MPC is providing the solution where the predicted quantiles satisfy the constraints. Since the learned quantiles have already captured the possible state distribution under disturbance, the proposed robust MPC consequently allows future states to deviate from nominal values while still satisfying constraints.

\begin{figure}[h!]
    \centering
    \includegraphics[width=0.6\linewidth]{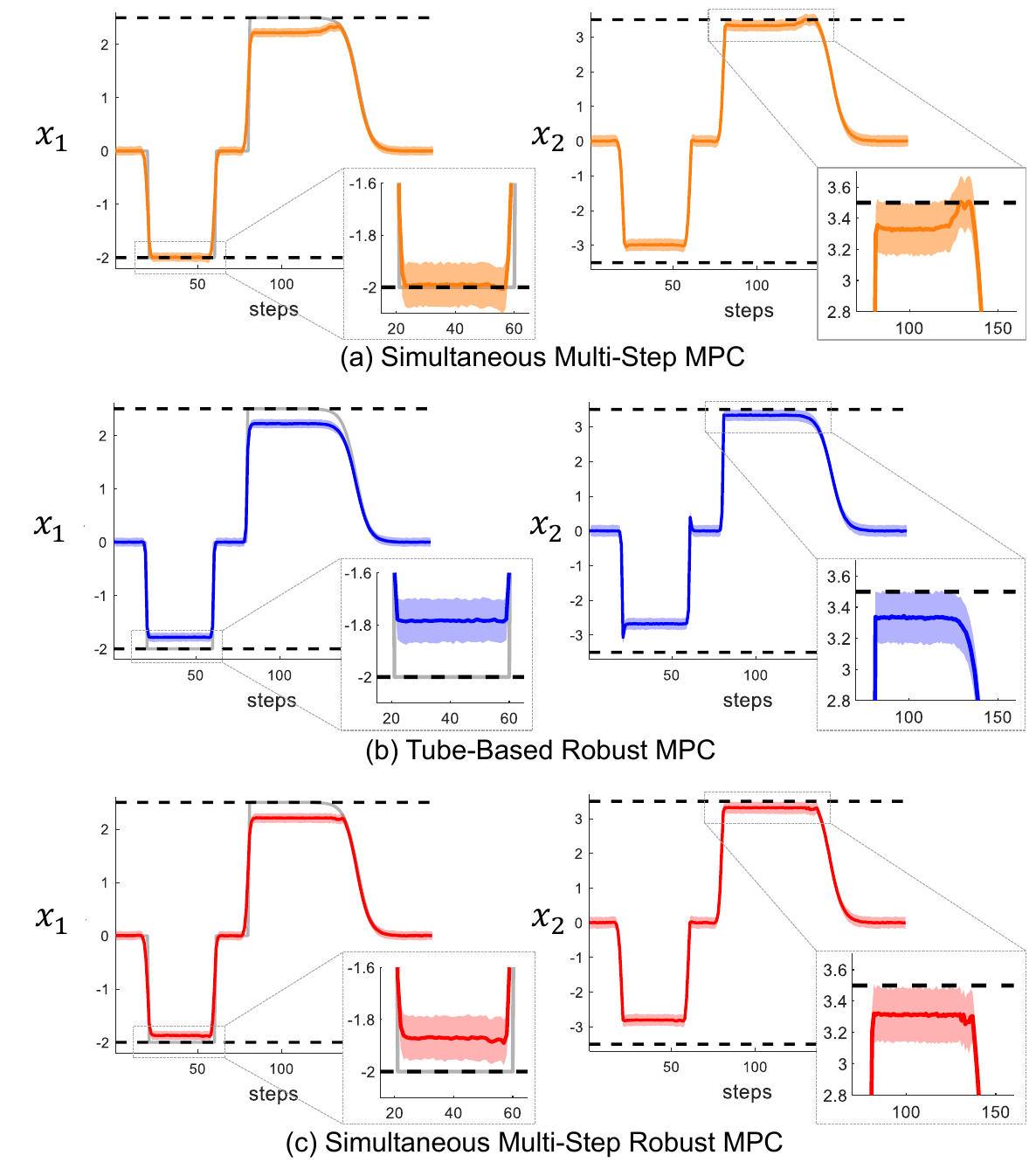}
    \caption{Distribution of trajectories under 1000 replicates. (a) Trajectories of multi-step (nominal) MPC. (b) Trajectories of tube-based robust MPC. (c) Trajectories of the proposed robust MPC.}
    \label{fig:toy_replicates}
\end{figure}

Finally, we compare the distribution of the output trajectories with tube-based MPC as the benchmark using 1,000 replicates. Tube-based MPC \cite{tsai2023robust} is a well-established, computationally efficient, and widely understood and applied approach that provides robust constraint satisfaction, particularly effective when the system is linear. However, since the error bound in tube-based MPC is approximated using the worst-case scenario, it may be too conservative in practical applications. Figure \ref{fig:toy_replicates} illustrates the distribution of the trajectories, where the thick lines represent the median of the trajectories at each time instance, and the color shades indicate the interval between 0.05 and 0.95 quantiles. In this study, we examine the 1,000 replicates at each timestep to calculate the constraint violation rate. We then use the maximum rate among the entire trajectory to represent the failure rate of each method. Figure \ref{fig:toy_replicates}(a) shows that without robust consideration, MPC easily violates the constraint, resulting in a 56.3\% failure rate. In Figure \ref{fig:toy_replicates}(b), although tube-based MPC yields 6.2\% failure rate and exhibits reliable performance, the margin between the lower bound of $x_1$ and the reference/constraint is significant, echoing the over-conservative feature. On the other hand, in Figure \ref{fig:toy_replicates}(c), the proposed robust MPC achieves 5.8\% failure rate and exhibits a smaller margin compared to that from the tube-based MPC. These results suggest that the learned quantile can be utilized as UQ while performing robust MPC. Furthermore, since TiDE directly learns the multi-step ahead response distribution from the data rather than approximating the error bound through uncertainty propagation, the simultaneous multi-step 
 robust MPC exhibits less conservative uncertainty estimation, leading to improved performance.

\section{Engineering Case study: Directed Energy Deposition Additive Manufacturing}


In this section, we implement the proposed multi-step robust MPC as the online decision-making process for the Digital Twin of the Directed Energy Deposition (DED) AM system. Given the inherent uncertainty associated with material variability and environmental factors in the DED process, proactive control strategies, such as MPC, become crucial to achieve desired material properties while minimizing defects \cite{gunasegaram2024machine}. Furthermore, the intricate dynamics of the melt pool makes it challenging to develop a physics-based model capable of providing accurate predictions in real-time. Therefore, data-driven methods have become promising tools for addressing this challenge. 

In this case study, our Digital Twin focuses exclusively on the melt pool rather than the entire part to enable real-time decision-making for process control. The melt pool is the most dynamic and sensitive region of the process, where critical quality indicators such as porosity, microstructure, and residual stress originate. Modeling the entire part with sufficient fidelity for in-situ control would be computationally prohibitive and incompatible with the actionable time requirements of online decision-making. In contrast, a melt pool-level Digital Twin allows for high-resolution, low-latency predictions that align with the frequency and spatial resolution of available in-situ sensing data, such as pyrometer readings. Moreover, control inputs like laser power directly influence melt pool behavior, making it the most actionable domain for decision-making via MPC. By concentrating modeling efforts on the melt pool, the Digital Twin remains both computationally tractable and operationally relevant, enabling accurate state prediction and MPC throughout the build process.

The Digital Twin of the DED process in this work comprises three key modules: (a) the physical system, represented by a high-fidelity finite element analysis (FEA) simulation, (b) the virtual system, implemented as a TiDE model that predicts melt pool temperature and depth, and (c) the virtual-to-physical integration, achieved through the proposed robust MPC framework. At each MPC timestep, the current melt pool temperature and depth are extracted from the physical system and used as \emph{feedback} to the virtual system. The virtual system then predicts future melt pool behavior conditioned on both the feedback and the candidate control input, which is subsequently optimized by the robust MPC. Once the optimal control input is determined, it is \emph{applied} to the physical system, completing the bi-directional interaction between the physical and virtual domains.

\subsection{Physical System: DED Setup}

\begin{figure}[ht]
  \centering
  \begin{minipage}{0.48\linewidth}  
    \centering
    \includegraphics[width=0.7\linewidth]{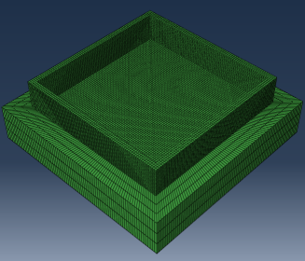}
    \caption{Single-track square}
    \label{fig:printed_square}
  \end{minipage}%
  \hfill
  \begin{minipage}{0.5\linewidth}  
    \centering
    \captionsetup{type=table}  
    \caption{Specification of the printed square} 
    \label{tab:square_spec}
    \footnotesize
    \begin{tabular}{c|c}
    \hline \hline
    Item             & Quantity  \\ \hline
    Side length      & 40 mm     \\
    Track width      & 1.5 mm    \\
    Layer height     & 0.75 mm   \\
    Num. of layers   & 10 layers \\
    Element size     & 0.375mm   \\
    Num. of elements & 40540     \\
    Substrate height & 10 mm       \\
    Scanning speed   & 7 mm/sec  \\ \hline \hline
    
    \end{tabular}
  \end{minipage}
\end{figure}

In this study, the physical DED is replaced by an in-house developed explicit FEA code developed by Liao \cite{liao2023efficient}. The code is accelerated by GPU computation using CuPy. It is employed for part-scale transient heat transfer simulation of the DED process. We select a single-track square as the target geometry, as shown in Fig. \ref{fig:printed_square}, and its specifications are listed in Table. \ref{tab:square_spec}. This numerical setup allows efficient simulation of complex thermal dynamics while maintaining high accuracy in capturing melt pool behavior. Here, we highlight that since the layer height is 0.75mm, we can set $x_{\text{depth,lb}}$ and $x_{\text{depth,ub}}$ to 0.225 and 0.075, corresponding to 10\% and 30\% dilution. For the details of the setup of DED, feature extraction, and data processing, please refer to our previous work \cite{CHEN2025412} for the technical details.

\subsection{Virtual System: Surrogate Modeling for the Melt Pool of DED}

To effectively generate a variety of laser power profile, the design of experiment (DOE) of the time series of the laser power profile is implemented. This method, proposed by Karkaria \cite{karkaria2024towards}, represents each time series with a 10-dimensional space using Fourier transform, and generates laser power trajectories using Latin Hypercube Sampling. For details of this approach, please refer to our previous work \cite{karkaria2024towards, CHEN2025412}. A total of 100 simulations with varying laser power profiles are conducted. The melt pool temperature and depth are extracted from the FEA model at each timestep. This diverse dataset ensures that TiDE can learn the relationship between process parameters and thermal responses across a wide operational range. As the data are generated, we train a TiDE model as a multi-variate multi-step ahead predictor with window size $w=50$ and horizon size $N=50$, and the 0.95 and 0.05 quantile are assigned as the upper and lower bounds. The specifics of training details, as well as model evaluation, are provided in our previous work \cite{CHEN2025412}. As a result, the MAPE and the RRMSE for melt pool temperature predictions are 1.29\% and 0.054, respectively, and those for the melt pool depth are 4.25\% and 0.0441. 

In our simulation, the primary sources of data uncertainty are the numerical errors in the FEA simulation. Specifically, since the laser’s travel distance at each simulation timestep does not correspond to the element size, the heat treatment time of each element will vary, resulting in substantial fluctuations in both melt pool temperature and depth. Consequently, these fluctuations will overshadow the effects of injected uncertainties on material variability or disturbances of our quantities of interest. This variability introduces aleatoric uncertainty, which TiDE effectively captures through its quantile-based predictions. Although this type of noise is repeatable, it appears to be irreducible by increasing data collection, as evidenced by the TiDE prediction. In fact, TiDE only extracts the pertinent features and smooths the nominal (median) response using the dense encoder. It then allows the learning quantile to handle the fluctuations within the training set. Given the objective of this work to demonstrate how the predicted model quantifies uncertainty and leverages it to enhance decision-making in Digital Twins, we contend that this source of uncertainty presents a more extreme scenario to evaluate the efficacy of the proposed method.

\subsection{Virtual-to-Physical Integration: Robust MPC for DED}
 The objective of implementing MPC in DED is to establish a proactive control strategy that effectively mitigates defects when an arbitrary reference trajectory for melt pool temperature is provided. In DED, porosity emerges as the most prevalent and critical defect, directly impacting the mechanical properties of printed components. Therefore, to mitigate defects, it is suggested to maintain the melt pool depth within a dilution range of 10\% and 30\% to avoid interlayer and intralayer porosity \cite{dass2019state}. Here, we assume that the melt pool depth is observable. In our previous work \cite{CHEN2025412}, the simultaneous multi-step MPC has been successfully implemented in melt pool depth constraint handling, using only nominal MPC. However, due to the intrinsic aleatoric uncertainty in the collected data and the processing environment, we aim to extend our previous work to perform robust MPC to enhance the constraint satisfaction rate. 

The robust MPC for melt pool temperature tracking can be formulated as:
\begin{subequations}
    
\begin{align}
 \min\limits_{\mathbf{u}} & \sum_{i=0}^{N-1} \left [ ||\hat{x}_{\text{temp},k+i+1} - r_{\text{temp},k+i+1}||^2_{\mathbf{Q}}  + ||\Delta u_{k+i}||^2_{\mathbf{R}}\right ]  \\
\text{s.t.} \: & \text{Pr}\left(\hat{x}_{\text{depth},k+i} \geq x_{\text{depth,lb}}\right)\geq0.95, \: ~\forall i\in\mathbb{N}_{[1,N]},   \\
 \quad & \text{Pr}\left(\hat{x}_{\text{depth},k+i} \leq x_{\text{depth,ub}}\right)\geq0.95, \: ~\forall i\in\mathbb{N}_{[1,N]},  \\
& [\hat{\mathbf{x}}_{\text{temp},k+1}^{f}, \hat{\mathbf{x}}_{\text{depth},k+1}^{f}]^T = \text{TiDE}(\mathbf{x}_{\text{temp},k}^{p}, \mathbf{x}_{\text{depth},k}^{p}, \mathbf{d}_{x,k}^{p:f},{d}_{y,k}^{p:f}, \mathbf{z}_k^{p:f},\mathbf{u}_k^{p:f}), \\
& {u}_{k+i} \in \mathbb{U}:=\{ u\in\mathbb{R}~|~ 504~W \leq u_i \leq 750~W\},
\end{align} 
\end{subequations}

\noindent where $\Delta u_{k+i-1}$ represents the differences between two consecutive terms in the designed future laser power, the distance between the laser nozzle and the closest edge on the $x$- and $y$-directions are denoted as $\mathbf{d}_x$ and $\mathbf{d}_y$, respectively. $\mathbf{z}$ represents the nozzle location on the $z$-direction. These three quantities are determined based on the geometry and are treated as additional covariates to enhance the prediction capabilities of TiDE. In this case, we do not tighten the input bound since the bounds will not be active throughout the process.

\subsection{MPC Results}
\begin{figure}[t]
    \centering
    \includegraphics[width=1\linewidth]{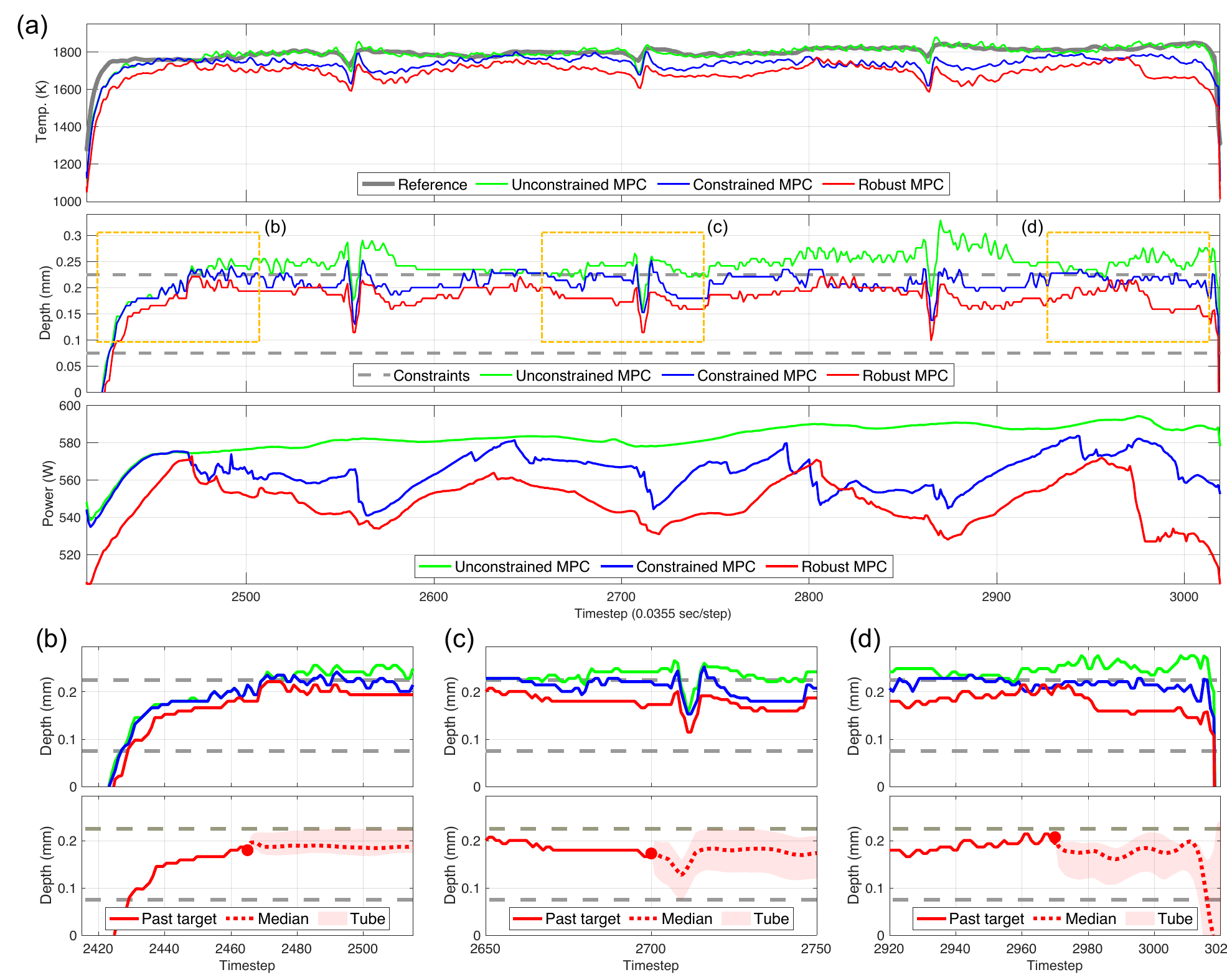}
    \caption{DED result comparison. (a) Trajectories of melt pool temperature, melt pool depth, and the corresponding laser power input. (b), (c), and (d) are the selected highlights comparing the details of the trajectories from different MPC methods, as well as the predicted medians and tubes at particular timesteps (red dots).}
    \label{fig:DED}
\end{figure}

The outcomes of implementing a robust MPC algorithm in DED are presented in Figure \ref{fig:DED}, where they are compared with the results obtained using unconstrained MPC and constrained MPC (which employs nominal prediction only), all employing the identical TiDE model as the multi-step ahead predictor. For a more detailed analysis, we have selected the trajectory from layer 5 as the representative. Figure \ref{fig:DED} illustrates the trajectory across the entire layer, encompassing the start of a new layer, the abrupt temperature fluctuations at the three corners due to the sharp change in scanning velocity, and the termination of the layer where the laser power is turned off. Given that the temperature/depth jumps and drops at the corners are unavoidable, we disregard the constraint violation penalty within a radius of 1 mm centered on the turning point. 

Figure \ref{fig:DED}(a) presents a comparison of the trajectory of melt pool temperature and depth, along with the corresponding laser power input. The unconstrained MPC demonstrates exceptional reference tracking performance, with its trajectory yielding $r^2=0.9730$ (for the entire trajectory) compared to the reference. This affirms the effectiveness of employing TiDE as a surrogate for multi-step MPC. However, as constraints are enforced, the constrained MPC compromises its reference tracking performance in favor of constraint satisfaction, where $r^2$ drops to 0.8261. As depicted in the figure, the resulting melt pool depth precisely adheres to the upper bound of the depth constraint. Nevertheless, since only nominal predictions are utilized in MPC and a safety buffer is not established, the constrained MPC occasionally violates the depth constraint. In contrast, robust MPC takes into account the potential state distribution, thereby generating a larger safety buffer from the melt pool depth constraint. Consequently, it effectively mitigates the constraint violation rate by compromising more on reference tracking, which results in $r^2=0.6920$. 

Figures \ref{fig:DED}(b) to \ref{fig:DED}(d) highlight the critical regions on melt pool depth that worth close examination. The upper subfigures zoom in on the comparison of the trajectories, while the lower subfigures present the predicted median and tube resulting from the optimal control inputs solved by robust MPC. Figure \ref{fig:DED}(b) illustrates the rise at the beginning of the layer, where the trajectory of the constrained MPC violates the constraint. It is evident that the upper bound of the tube is utilized in robust MPC to adjust from the constraint, accommodating the disturbance during the process. Figure \ref{fig:DED}(c) exhibits that TiDE, along with its predicted tube, captures the distribution of depth variation at the corner. Figure \ref{fig:DED}(d) illustrates that since the laser power will be deactivated, and the feasible solution is not attainable from timestep 2960 until the end of the layer because to the lower bound cannot be satisfied, the lower bound constraint is relaxed at this region to ensure feasibility.

\subsection{Computational Time}

\begin{figure}[h!]
    \centering
    \includegraphics[width=0.5\linewidth]{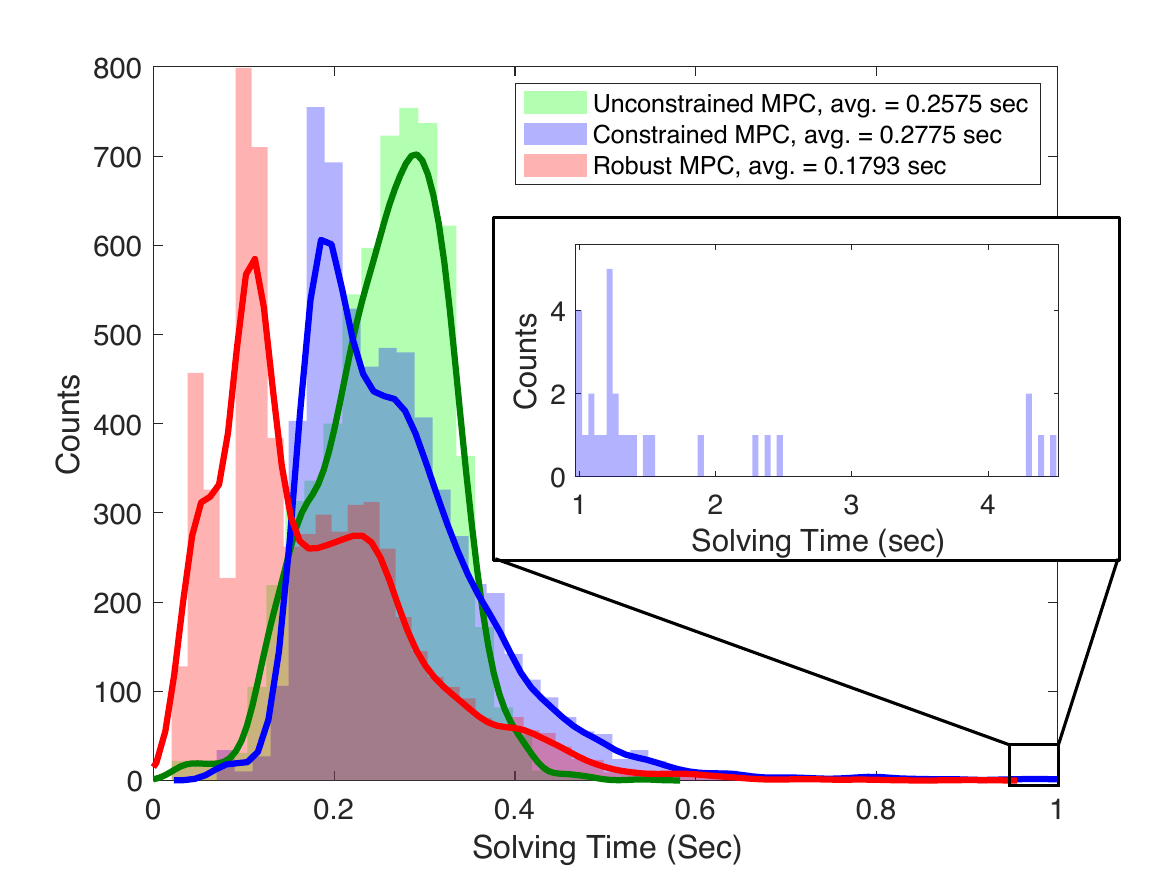}
    \caption{Histogram of MPC solving time.}
    \label{fig:time_hist}
\end{figure}

The histogram presented in Figure \ref{fig:time_hist} illustrates the computational time required to solve the unconstrained, constrained, and robust MPC problems at each step, computed using an AMD Ryzen Threadripper PRO 3975WX 32-Cores CPU. The results demonstrate that the average computational time for robust MPC is 0.1793 seconds, with a maximum of 0.903 seconds. These findings indicate that the proposed method can be effectively applied in various real-world scenarios, enabling real-time decision-making for Digital Twins.
The relatively faster solving time of robust MPC compared to unconstrained MPC can be attributed to the creation of a safety buffer by the tube constraint. This buffer limits the feasible solution space, potentially leading to a shorter optimization process. Furthermore, robust MPC also results in a shorter solution time compared to constrained MPC. The primary reason for this is that constrained MPC encounters more constraint violations, requiring more iterations to converge. This is because employing an infeasible solution as the initial guess for the penalty method hinders its constraint-handling capability.  

This case study verifies the suitability of our method for complex engineering applications. It highlights TiDE’s capability as a surrogate model for multi-step predictions with UQ and demonstrates the effectiveness of multi-step robust MPC in efficiently handling constraints and solving MPC problems. The integration of these approaches presents a robust and practical framework for decision-making processes of the Digital Twin of engineering systems.

\section{Closure}

This work proposed a simultaneous multi-step robust MPC framework that integrates TiDE with quantile regression to enable real-time decision-making for Digital Twins with uncertainty awareness. By leveraging TiDE's capacity for multi-step predictions and efficient UQ from quantile regression, the proposed framework demonstrated an effective approach to quantify aleatoric uncertainty, and further benefit solving robust MPC with a series of acceleration techniques using automatic differentiation. In contrast to conventional single-step MPC approaches that necessitate recursive roll-out for state prediction and conservative uncertainty approximation, the simultaneous multi-step predictions reduced the number of function calls associated with recursive state propagation. Furthermore, the quantile-based uncertainty representation improved constraint satisfaction in the presence of stochastic disturbances. Through the validation of numerical simulations and engineering case studies employing DED, we demonstrate the exceptional surrogate modeling capabilities of TiDE for complex system dynamics with multi-step ahead prediction. Furthermore, we highlight the potential of this learning-based MPC framework to provide precise and proactive control strategies for intricate, nonlinear systems. This establishes a foundation for future advancements in uncertainty-aware Digital Twin applications.

While this work demonstrates substantial advancements, several limitations remain. To begin with, the effectiveness of the proposed framework relies heavily on the quality and diversity of the training data, which may limit its generalizability to scenarios involving unseen disturbances or operating conditions not captured during model training. Additionally, while TiDE reduces computational overhead compared to traditional methods, the computational demands may still pose challenges for real-time applications with high-dimensional design space. Furthermore, the current implementation lacks mechanisms for dynamic adaptation to evolving system dynamics or disturbances beyond pre-trained models, which could limit its robustness in highly variable environments. Moreover, we only demonstrate the effectiveness of the proposed method on stable systems, while it might be challenging for the implementation of unstable systems. Lastly, this study does not carry out comprehensive proofs of stability, recursive feasibility, and performance guarantee, convergence, etc. These limitations highlight opportunities for future work to enhance the framework's adaptability, efficiency, and applicability across more complex and unpredictable systems.

In the future, we will develop a framework that enables the dynamic adaptation of the surrogate model through effective parameter fine-tuning methods. This will enhance the resilience, trustworthiness, and flexibility of the surrogate model as well as the decision-making process, thereby fulfilling the full potential of Digital Twin systems.

\section*{Acknowledgement}
We are grateful for the grant support from the National Science Foundation’s Engineering Research Center for Hybrid Autonomous Manufacturing: Moving from Evolution to Revolution (ERC-HAMMER), under the Award Number EEC-2133630. Yi-Ping Chen also acknowledges the Taiwan-Northwestern Doctoral Scholarship and the fellowship support from Northwestern University for the Predictive Science \& Engineering Design (PS\&ED) Cluster project. We acknowledge the technical support from Faith Rolark and Daniel Quispe on the DED simulation, and the insightful discussion with Dr. Aaditya Chandrasekhar, Dr. Jian Cao, and Dr. Robert X. Gao.

\addcontentsline{toc}{section}{Reference}
\bibliographystyle{asmems4}
\bibliography{main}

\titleformat{\section}[hang]{\fontsize{11pt}{0pt}\selectfont\bf}{\thesection}{0cm}{}

\titleformat{\section}[hang]{\centering\fontsize{11pt}{0pt}\selectfont\bf}{}{0.1cm}{}

\end{document}